# Octonic relativistic quantum mechanics

V.L.Mironov, S.V.Mironov

Institute for physics of microstructures RAS, 603950, Nizhniy Novgorod, GSP-105, Russia
e-mail: mironov@ipm.sci-nnov.ru

In this paper we represent the generalization of relativistic quantum mechanics on the base of eght-component values "octons", generating associative noncommutative spatial algebra. It is shown that the octonic second-order equation for the eight-component octonic wave function, obtained from the Einshtein relation for energy and momentum, describes particles with spin of 1/2. It is established that the octonic wave function of a particle in the state with defined spin projection has the specific spatial structure in the form of octonic oscillator with two spatial polarizations: longitudinal linear and transversal circular. The relations between bispinor and octonic descriptions of relativistic particles are established.

We propose the eight-component spinors, which are octonic generalisation of two-component Pauli spinors and four-component Dirac bispinors. It is shown that proposed eight-component spinors separate the states with different spin projection, different particle-antiparticle state as well as different polarization of the octonic oscillator.

We demonstrate that in the frames of octonic relativistic quantum mechanics the second-order equation for octonic wave function can be reformulated in the form of the system of first-order equations for quantum fields, which is analogous to the system of Maxwell equations for the electromagnetic field. It is established that for the special type of wave functions the second-order equation can be reduced to the single first-order equation, which is analogous to the Dirac equation. At the same time it is shown that this first-order equation describes particles, which do not create quantum fields.

**Contents:**

## 1. Introduction

Multi-component hypernumbers [1-5] are widely used in relativistic mechanics, electrodynamics, quantum mechanics and quantum field theory [6-23]. For instance the structure of four-component quaternions adequately conforms to the four-dimensional character of space-time and allows to develop quaternionic generalizations of quantum mechanics [6-12], where particles are described by four-component wave function consisting of scalar and vector parts. However quaternionic wave functions have no pseudoscalar and pseudovector properties and are transformed incorrectly in respect to the spatial inversion. Therefore it seams that eight-component values including scalar, pseudoscalar, vector and pseudovector components are more appropriate for description of relativistic quantum systems.

However the attempts to describe relativistic particles by means of different eight-component hypernumbers such as biquaternions [5, 15, 16], octonions [17-20] and multi-vectors generating associative Clifford algebras [21-23] have no any appreciable progress. For example, few attempts to describe the relativistic particles by means of octonion wave functions are confronted by difficulties, which are connected with octonions nonassociativity [19]. Moreover all systems of hypercomplex numbers, which have been applied up to now for quantum mechanics generalization (quaternions, biquaternions, octonions and multivectors) are the objects of hypercomplex space and have not any consistent space-geometric interpretation.

Recently we proposed eight-component values "octons" generating closed noncommutative associative algebra and having clear well-defined geometric interpretation [24]. From geometrical point of view octon is the object of the real three-dimensional space. It is the sum of scalar, pseudoscalar, vector and pseudovector. In [24] octons were successfully applied for the classical electromagnetic field description. In present paper we propose the generalization of relativistic quantum mechanics on the base of octonic equations for eight-component octonic wave functions.

## 2. Algebra of octons

In the beginning we will shortly remind the basic properties of octons. The eight-component octon $\breve{G}$ is defined in the form

$$\breve{G} = d + a\boldsymbol{i} + b\boldsymbol{j} + c\boldsymbol{k} + D\boldsymbol{E} + A\boldsymbol{I} + B\boldsymbol{J} + C\boldsymbol{K}, \tag{1}$$

where values $\boldsymbol{i}, \boldsymbol{j}, \boldsymbol{k}$ are polar unit vectors; $\boldsymbol{I}, \boldsymbol{J}, \boldsymbol{K}$ are axial unit vectors and $\boldsymbol{E}$ is the pseudoscalar unit. The octon's components $d, a, b, c, D, A, B, C$ are the numbers (complex in general). Thus the octon is the sum of scalar, vector, pseudoscalar and pseudovector and the $\boldsymbol{i}, \boldsymbol{j}, \boldsymbol{k}, \boldsymbol{I}, \boldsymbol{J}, \boldsymbol{K}$ values are polar and axial bases of octon respectively. The algebra of octons was discussed in detail in [24]. Here we shortly remind of the basic commutation and multiplication rules which are represented in the table 1.

*Table 1. The rules of multiplication and commutation for the octon's unit vectors.*

|   | $\boldsymbol{i}$ | $\boldsymbol{j}$ | $\boldsymbol{k}$ | $\boldsymbol{E}$ | $\boldsymbol{I}$ | $\boldsymbol{J}$ | $\boldsymbol{K}$ |
|---|---|---|---|---|---|---|---|
| $\boldsymbol{i}$ | $1$ | $\xi\boldsymbol{K}$ | $-\xi\boldsymbol{J}$ | $\boldsymbol{I}$ | $\boldsymbol{E}$ | $\xi\boldsymbol{k}$ | $-\xi\boldsymbol{j}$ |
| $\boldsymbol{j}$ | $-\xi\boldsymbol{K}$ | $1$ | $\xi\boldsymbol{I}$ | $\boldsymbol{J}$ | $-\xi\boldsymbol{k}$ | $\boldsymbol{E}$ | $\xi\boldsymbol{i}$ |
| $\boldsymbol{k}$ | $\xi\boldsymbol{J}$ | $-\xi\boldsymbol{I}$ | $1$ | $\boldsymbol{K}$ | $\xi\boldsymbol{j}$ | $-\xi\boldsymbol{i}$ | $\boldsymbol{E}$ |
| $\boldsymbol{E}$ | $\boldsymbol{I}$ | $\boldsymbol{J}$ | $\boldsymbol{K}$ | $1$ | $\boldsymbol{i}$ | $\boldsymbol{j}$ | $\boldsymbol{k}$ |
| $\boldsymbol{I}$ | $\boldsymbol{E}$ | $\xi\boldsymbol{k}$ | $-\xi\boldsymbol{j}$ | $\boldsymbol{i}$ | $1$ | $\xi\boldsymbol{K}$ | $-\xi\boldsymbol{J}$ |
| $\boldsymbol{J}$ | $-\xi\boldsymbol{k}$ | $\boldsymbol{E}$ | $\xi\boldsymbol{i}$ | $\boldsymbol{j}$ | $-\xi\boldsymbol{K}$ | $1$ | $\xi\boldsymbol{I}$ |
| $\boldsymbol{K}$ | $\xi\boldsymbol{j}$ | $-\xi\boldsymbol{i}$ | $\boldsymbol{E}$ | $\boldsymbol{k}$ | $\xi\boldsymbol{J}$ | $-\xi\boldsymbol{I}$ | $1$ |

In the table 1 and further the value $\xi$ is the imaginary unit: $\xi^2 = -1$.

## 3. Octonic wave function and spatial operators

Let us consider the wave function of the relativistic particle in the form of eight-component octon

$$\breve{\psi} = \psi_0 + \psi_1 \boldsymbol{i} + \psi_2 \boldsymbol{j} + \psi_3 \boldsymbol{k} + \varphi_0 \boldsymbol{E} + \varphi_1 \boldsymbol{I} + \varphi_2 \boldsymbol{J} + \varphi_3 \boldsymbol{K} .\qquad(2)$$

The components $\psi_\alpha(\vec{r},t)$ and $\varphi_\alpha(\vec{r},t)$ ($\alpha = 0,1,2,3$) are scalar (complex in general) functions of spatial coordinates and time. The octonic wave function (2) can be written also in compact form

$$\breve{\psi} = \psi_0 + \vec{\psi} + \tilde{\varphi}_0 + \vec{\varphi} ,\qquad(3)$$

where vector part is indicated by arrow "→", pseudoscalar part by the wave " ~ ", and pseudovector part by double arrow "↔".

The wave function of the free particle should satisfy an equation, which is obtained from the Einstein relation between particle energy and momentum

$$E^2 - p^2 c^2 = m^2 c^4 \qquad(4)$$

by means of changing classical momentum $\vec{p}$ and energy $E$ on corresponding quantum-mechanical operators $\hat{\vec{p}} = -\xi\hbar\vec{\nabla}$ and $\hat{E} = \xi\hbar\dfrac{\partial}{\partial t}$. This equation has the following form:

$$\left(\frac{1}{c^2}\frac{\partial^2}{\partial t^2} - \Delta\right)\breve{\psi} = -\frac{m^2 c^2}{\hbar^2}\breve{\psi} .\qquad(5)$$

Here $c$ is the velocity of light, $m$ is the mass of the particle, $\hbar$ is the Plank constant. In contrast to the scalar Klein-Gordon equation [25] the expression (5) is the octonic equation since it is written for the octonic function. It is clear that each of $\breve{\psi}$ components satisfies the scalar Klein-Gordon equation.

The operator in the left part of equation (5) can be represented as the product of two operators:

$$\left(\frac{1}{c}\frac{\partial}{\partial t} - \vec{\nabla}\right)\left(\frac{1}{c}\frac{\partial}{\partial t} + \vec{\nabla}\right)\breve{\psi} = -\frac{m^2 c^2}{\hbar^2}\breve{\psi} .\qquad(6)$$

Here we assume that the octonic wave function $\breve{\psi}$ is twice continuously differentiable, so $[\vec{\nabla},\vec{\nabla}]\breve{\psi} = 0$. The equation (6) can be represented in expanded form

$$\left(\frac{1}{c}\frac{\partial}{\partial t} - \frac{\partial}{\partial x}\boldsymbol{i} - \frac{\partial}{\partial y}\boldsymbol{j} - \frac{\partial}{\partial z}\boldsymbol{k}\right)\left(\frac{1}{c}\frac{\partial}{\partial t} + \frac{\partial}{\partial x}\boldsymbol{i} + \frac{\partial}{\partial y}\boldsymbol{j} + \frac{\partial}{\partial z}\boldsymbol{k}\right)\breve{\psi}(\vec{r},t) = -\frac{m^2 c^2}{\hbar^2}\breve{\psi}(\vec{r},t),\qquad(7)$$

where the differential operators force on time and spatial coordinates of the wave function $\breve{\psi}(\vec{r},t)$. Note that unit vectors $\boldsymbol{i}, \boldsymbol{j}, \boldsymbol{k}$ in the left part of equation (7) transform the spatial structure of the wave function by means of octonic multiplication. In this sense they can be treated as spatial operators $\hat{\boldsymbol{i}}, \hat{\boldsymbol{j}}, \hat{\boldsymbol{k}}$, which transform octon of the wave function. For example the action of $\hat{\boldsymbol{i}}$ operator can be represented as octonic multiplication of unit vector $\boldsymbol{i}$ and octon $\breve{\psi}$:

$$\hat{\boldsymbol{i}}\breve{\psi} = \boldsymbol{i}\breve{\psi} = \psi_1 + \psi_0 \boldsymbol{i} - \xi\varphi_3 \boldsymbol{j} + \xi\varphi_2 \boldsymbol{k} + \varphi_1 \boldsymbol{E} + \varphi_0 \boldsymbol{I} - \xi\psi_3 \boldsymbol{J} + \xi\psi_2 \boldsymbol{K} .\qquad(8)$$

Further we will use symbolic designations $\hat{\boldsymbol{i}}, \hat{\boldsymbol{j}}, \hat{\boldsymbol{k}}$ in the operator parts of equations but $\boldsymbol{i}, \boldsymbol{j}, \boldsymbol{k}$ designations in the wave functions. Then the equation (7) can be rewritten in the operator form

$$\left(\frac{1}{c}\frac{\partial}{\partial t} - \frac{\partial}{\partial x}\hat{\boldsymbol{i}} - \frac{\partial}{\partial y}\hat{\boldsymbol{j}} - \frac{\partial}{\partial z}\hat{\boldsymbol{k}}\right)\left(\frac{1}{c}\frac{\partial}{\partial t} + \frac{\partial}{\partial x}\hat{\boldsymbol{i}} + \frac{\partial}{\partial y}\hat{\boldsymbol{j}} + \frac{\partial}{\partial z}\hat{\boldsymbol{k}}\right)\breve{\psi}(\vec{r},t) = -\frac{m^2 c^2}{\hbar^2}\breve{\psi}(\vec{r},t) .\qquad(9)$$

On the other hand the expression (8) can be represented in equivalent matrix form as the action of the matrix operator $\hat{i}$ on the eight-component column of the wave function, so $\hat{i}$ can be written as the $8 \times 8$ matrix:

$$\hat{i} \begin{pmatrix} \psi_0 \\ \psi_1 \\ \psi_2 \\ \psi_3 \\ \varphi_0 \\ \varphi_1 \\ \varphi_2 \\ \varphi_3 \end{pmatrix} = \begin{pmatrix} \psi_1 \\ \psi_0 \\ -\xi\varphi_3 \\ \xi\varphi_2 \\ \varphi_1 \\ \varphi_0 \\ -\xi\psi_3 \\ \xi\psi_2 \end{pmatrix} \Rightarrow \hat{i} = \begin{pmatrix} 0 & 1 & 0 & 0 & 0 & 0 & 0 & 0 \\ 1 & 0 & 0 & 0 & 0 & 0 & 0 & 0 \\ 0 & 0 & 0 & 0 & 0 & 0 & 0 & -\xi \\ 0 & 0 & 0 & 0 & 0 & 0 & \xi & 0 \\ 0 & 0 & 0 & 0 & 0 & 1 & 0 & 0 \\ 0 & 0 & 0 & 0 & 1 & 0 & 0 & 0 \\ 0 & 0 & 0 & -\xi & 0 & 0 & 0 & 0 \\ 0 & 0 & \xi & 0 & 0 & 0 & 0 & 0 \end{pmatrix}. \quad (10)$$

The matrix representation for the rest operators of octon's basis $\hat{j}$, $\hat{k}$, $\hat{I}$, $\hat{J}$, $\hat{K}$, $\hat{E}$ is given in the appendix 1. Further we will use vector designation for the operators of octon's basis since it is more compact.

Thus in octonic quantum mechanics the spatial operators $\hat{i}$, $\hat{j}$, $\hat{k}$, $\hat{I}$, $\hat{J}$, $\hat{K}$, $\hat{E}$, which transpose components of the wave function, are used parallel with differentiation operators.

Let us define the operation of spatial inversion ($R$) of octonic wave function. This operation reverses the vector component of the wave function and changes the sign of the pseudoscalar component. In particularly the simplest octonic wave functions consisting only of one element of octon's basis are transformed under spatial inversion in accordance with the following rules:

$$\begin{aligned} R : \boldsymbol{i}, \boldsymbol{j}, \boldsymbol{k} &\Rightarrow -\boldsymbol{i}, -\boldsymbol{j}, -\boldsymbol{k} \\ R : \boldsymbol{I}, \boldsymbol{J}, \boldsymbol{K} &\Rightarrow \boldsymbol{I}, \boldsymbol{J}, \boldsymbol{K} \\ R : \boldsymbol{E} &\Rightarrow -\boldsymbol{E} \end{aligned} \quad (11)$$

Note that

$$\boldsymbol{E} = -\xi\, \boldsymbol{i}\,\boldsymbol{j}\,\boldsymbol{k} \quad (12)$$

is the simplest pseudoscalar wave function which changes the sign under spatial inversion.

The operation of spatial inversion is realized by octonic operator $\hat{R}$ which changes the signs of vector and pseudoscalar components of the wave function:

$$\hat{R}\left(\psi_0 + \vec{\psi} + \tilde{\varphi}_0 + \vec{\varphi}\right) = \left(\psi_0 - \vec{\psi} - \tilde{\varphi}_0 + \vec{\varphi}\right). \quad (13)$$

We specially emphasize that the operator $\hat{R}$ does not act on arguments of the wave function i.e. for instance $\hat{R}\vec{\psi}(x,y,z,t) = -\vec{\psi}(x,y,z,t)$. The operator $\hat{R}$ has the following matrix representation:

$$\hat{R} = \begin{pmatrix} 1 & 0 & 0 & 0 & 0 & 0 & 0 & 0 \\ 0 & -1 & 0 & 0 & 0 & 0 & 0 & 0 \\ 0 & 0 & -1 & 0 & 0 & 0 & 0 & 0 \\ 0 & 0 & 0 & -1 & 0 & 0 & 0 & 0 \\ 0 & 0 & 0 & 0 & -1 & 0 & 0 & 0 \\ 0 & 0 & 0 & 0 & 0 & 1 & 0 & 0 \\ 0 & 0 & 0 & 0 & 0 & 0 & 1 & 0 \\ 0 & 0 & 0 & 0 & 0 & 0 & 0 & 1 \end{pmatrix}. \quad (14)$$

Following the definition (13), it is easy to show that the operator $\hat{R}$ anti-commutes with operators $\hat{i}$, $\hat{j}$, $\hat{k}$, $\hat{E}$ and commutes with operators $\hat{I}$, $\hat{J}$, $\hat{K}$.

Note that the action of $\hat{R}$ is defined only regarding the wave function (see (13)). Therefore for example the action of operator product $\hat{R}\hat{i}$ on the wave function can be represented only in two following forms:

$$\hat{R}\hat{i}\breve{\psi} = \hat{R}(\hat{i}\breve{\psi}) \text{ or} \qquad (15)$$

$$\hat{R}\hat{i}\breve{\psi} = -\hat{i}(\hat{R}\breve{\psi}). \qquad (16)$$

At that the operators $\hat{R}$ and $\hat{i}$ can not be multiplied:

$$\hat{R}\hat{i}\breve{\psi} \neq -\hat{i}\breve{\psi}. \qquad (17)$$

### 4. Eigenvalues and eigenfunctions of octonic spatial operators

Let us consider the eigenvalues and eigenfunctions of spatial operators by the example of the operator $\hat{K}$, which will be used below for the description of the particle in homogeneous magnetic field. The equation for the eigenvalues and eigenfunctions in this case has the following simple form:

$$\hat{K}\breve{\psi} = \lambda\breve{\psi}. \qquad (18)$$

Rewriting it in expanded form, we get

$$\hat{K}(\psi_0 + \psi_x \boldsymbol{i} + \psi_y \boldsymbol{j} + \psi_z \boldsymbol{k} + \varphi_0 \boldsymbol{E} + \varphi_x \boldsymbol{I} + \varphi_y \boldsymbol{J} + \varphi_z \boldsymbol{K}) = \lambda(\psi_0 + \psi_x \boldsymbol{i} + \psi_y \boldsymbol{j} + \psi_z \boldsymbol{k} + \varphi_0 \boldsymbol{E} + \varphi_x \boldsymbol{I} + \varphi_y \boldsymbol{J} + \varphi_z \boldsymbol{K}). \qquad (19)$$

Performing octonic multiplication in the left part of (19) we obtain

$$\psi_0 \boldsymbol{K} + \xi\psi_x \boldsymbol{j} - \xi\psi_y \boldsymbol{i} + \psi_z \boldsymbol{E} + \varphi_0 \boldsymbol{k} + \xi\varphi_x \boldsymbol{J} - \xi\varphi_y \boldsymbol{I} + \varphi_z = \lambda(\psi_0 + \psi_x \boldsymbol{i} + \psi_y \boldsymbol{j} + \psi_z \boldsymbol{k} + \varphi_0 \boldsymbol{E} + \varphi_x \boldsymbol{I} + \varphi_y \boldsymbol{J} + \varphi_z \boldsymbol{K}). \qquad (20)$$

Equating components in (20), we obtain the system of eight scalar equations

$$\begin{cases} \varphi_z = \lambda\psi_0, \\ -\xi\psi_y = \lambda\psi_x, \\ \xi\psi_x = \lambda\psi_y, \\ \varphi_0 = \lambda\psi_z, \\ \psi_z = \lambda\varphi_0, \\ -\xi\varphi_y = \lambda\varphi_x, \\ \xi\varphi_x = \lambda\varphi_y, \\ \psi_0 = \lambda\varphi_z. \end{cases} \qquad (21)$$

After trivial mathematical transformations the system (21) can be rewritten in the following form:

$$\begin{cases} \lambda^2 = 1, \\ \varphi_z = \lambda\psi_0, \\ \psi_y = \xi\lambda\psi_x, \\ \psi_z = \lambda\varphi_0, \\ \varphi_y = \xi\lambda\varphi_x. \end{cases} \qquad (22)$$

The first equation in (22) shows that $\lambda = \pm 1$. At that for each eigenvalue $\lambda$ there is the four-dimensional subspace of eigenfunctions. Taking into account (22) we can choose the set of functions

$$(1+\boldsymbol{K}),\ (\boldsymbol{i}+\xi\boldsymbol{j}),\ (\boldsymbol{E}+\boldsymbol{k}),\ (\boldsymbol{I}+\xi\boldsymbol{J}) \tag{23}$$

as the basis of the subspace corresponding to eigenvalue $\lambda = +1$, and set of functions

$$(1-\boldsymbol{K}),\ (\boldsymbol{i}-\xi\boldsymbol{j}),\ (\boldsymbol{E}-\boldsymbol{k}),\ (\boldsymbol{I}-\xi\boldsymbol{J}) \tag{24}$$

as the basis of the subspace corresponding to $\lambda = -1$.

Then arbitrary eigenfunctions of the operator $\hat{\boldsymbol{K}}$ corresponding to $\lambda = \pm 1$ can be represented in the form of linear combination of basis functions (23) or (24):

$$\breve{\psi}_{\lambda=1} = F_1^{(1)}(\vec{r},t)(1+\boldsymbol{K}) + F_2^{(1)}(\vec{r},t)(\boldsymbol{i}+\xi\boldsymbol{j}) + F_3^{(1)}(\vec{r},t)(\boldsymbol{E}+\boldsymbol{k}) + F_4^{(1)}(\vec{r},t)(\boldsymbol{I}+\xi\boldsymbol{J}), \tag{25}$$

$$\breve{\psi}_{\lambda=-1} = F_1^{(-1)}(\vec{r},t)(1-\boldsymbol{K}) + F_2^{(-1)}(\vec{r},t)(\boldsymbol{i}-\xi\boldsymbol{j}) + F_3^{(-1)}(\vec{r},t)(\boldsymbol{E}-\boldsymbol{k}) + F_4^{(-1)}(\vec{r},t)(\boldsymbol{I}-\xi\boldsymbol{J}), \tag{26}$$

where $F_\alpha^{(\lambda)}(\vec{r},t)$ are arbitrary scalar functions of space coordinates and time ($\alpha = 1, 2, 3, 4$; $\lambda = \pm 1$). It is obviously, that eigenvalues and eigenfunctions of the operator $\hat{\boldsymbol{K}}$ could be obtained also using matrix representation.

The eigenfunctions of other octonic spatial operators are represented in appendix 2. Note that the simplest eigenfunctions of the operator $\hat{\boldsymbol{K}}$ generate a specific closed algebra. The rules of multiplication of these functions are considered in the appendix 3.

**5. Octonic equation for relativistic particle in an external electromagnetic field**

To describe a particle in an external electromagnetic field the following change of quantum-mechanical operators in equations should be made [25]:

$$\hat{E} \to \hat{E} - e\Phi, \qquad \hat{\vec{p}} \to \hat{\vec{p}} - \frac{e}{c}\vec{A}, \tag{27}$$

where $\Phi$ and $\vec{A}$ are scalar and vector potentials of electromagnetic field, $e$ is the particle charge ($e < 0$ for the electron). The change (27) is equivalent to the following change of differential operators:

$$\frac{\partial}{\partial t} \to \frac{\partial}{\partial t} + \frac{\xi e}{\hbar}\Phi, \qquad \vec{\nabla} \to \vec{\nabla} - \frac{\xi e}{\hbar c}\vec{A}. \tag{28}$$

Substituting operators (28) in the equation (6) we obtain

$$\left(\frac{1}{c}\frac{\partial}{\partial t} + \frac{\xi e}{\hbar c}\Phi - \vec{\nabla} + \frac{\xi e}{\hbar c}\vec{A}\right)\left(\frac{1}{c}\frac{\partial}{\partial t} + \frac{\xi e}{\hbar c}\Phi + \vec{\nabla} - \frac{\xi e}{\hbar c}\vec{A}\right)\breve{\psi} = -\frac{m^2 c^2}{\hbar^2}\breve{\psi}. \tag{29}$$

The multiplication of octonic operators in the left part of (29) leads to the following equation:

$$\left[\frac{1}{c^2}\frac{\partial^2}{\partial t^2} - \Delta + \frac{2\xi e}{\hbar c}\left((\vec{A},\vec{\nabla}) + \frac{\Phi}{c}\frac{\partial}{\partial t}\right) + \frac{m^2 c^2}{\hbar^2} + \frac{e^2}{\hbar^2 c^2}(A^2 - \Phi^2)\right]\breve{\psi} - \frac{e}{\hbar c}\vec{H}\breve{\psi} + \frac{\xi e}{\hbar c}\vec{E}\breve{\psi} = 0. \tag{30}$$

Here we have taken into account that $\vec{E} = -\vec{\nabla}\Phi - \frac{1}{c}\frac{\partial \vec{A}}{\partial t}$ is the vector of the electric field, $\vec{H} = -\xi[\vec{\nabla},\vec{A}]$ is the pseudovector of the magnetic field, and $(\vec{\nabla},\vec{A}) + \frac{1}{c}\frac{\partial \Phi}{\partial t} = 0$ in view of the Lorentz gauge. The operations of scalar and vector octonic multiplication were considered in [24].

Note that the octonic equation (30) encloses the specific terms $-\frac{e}{\hbar c}\vec{H}\,\breve{\psi} + \frac{\xi e}{\hbar c}\vec{E}\,\breve{\psi}$, where the fields $\vec{E}$ and $\vec{H}$ play the role of spatial octonic operators.

## 6. Octonic equation for nonrelativistic particle in an external electromagnetic field

The octonic equation for nonrelativistic particle in an external electromagnetic field can be obtained directly on the base of the Schrödinger equation. The octonic Hamiltonian for the free particle has the form

$$\hat{H}_0 = \frac{\hat{\vec{p}}^{\,2}}{2m} = -\frac{\hbar^2}{2m}\Delta. \tag{31}$$

The octonic equation for the octonic wave function corresponding to Hamiltonian (31) can be written in the following form:

$$\xi\hbar\frac{\partial \breve{\psi}}{\partial t} = \hat{H}_0\breve{\psi}. \tag{32}$$

Without electromagnetic field this equation is equivalent to eight scalar Schrödinger equations for each component of the octonic wave function.

To describe a particle in electromagnetic field the change of quantum-mechanical operators $\hat{H}_0 \rightarrow \hat{H} = \hat{H}_0 + e\Phi$, $\vec{\nabla} \rightarrow \vec{\nabla} - \frac{\xi e}{\hbar c}\vec{A}$ should be done. In this case we obtain

$$\hat{H} = -\frac{\hbar^2}{2m}\left(\vec{\nabla} - \frac{\xi e}{\hbar c}\vec{A}\right)^2 + e\Phi. \tag{33}$$

Multiplying the octonic operators in (33) we get

$$\hat{H} = -\frac{\hbar^2}{2m}\Delta + \frac{\xi\hbar e}{2mc}(\vec{\nabla},\vec{A}) + \frac{\xi\hbar e}{mc}(\vec{A},\vec{\nabla}) + \frac{e^2}{2mc^2}A^2 + e\Phi - \frac{\hbar e}{2mc}\vec{H}. \tag{34}$$

Thus nonstationary octonic equation for nonrelativistic particle in electromagnetic field can be written as

$$\xi\hbar\frac{\partial \breve{\psi}}{\partial t} = \hat{H}\breve{\psi}. \tag{35}$$

In stationary state with energy $E$ the wave function satisfies the following equation:

$$-\frac{\hbar^2}{2m}\Delta\breve{\psi} + \frac{\xi\hbar e}{2mc}(\vec{\nabla},\vec{A})\breve{\psi} + \frac{\xi\hbar e}{mc}(\vec{A},\vec{\nabla})\breve{\psi} + \frac{e^2}{2mc^2}A^2\breve{\psi} + e\Phi\breve{\psi} - \frac{\hbar e}{2mc}\vec{H}\breve{\psi} = E\breve{\psi}. \tag{36}$$

Note that the last item $-\frac{\hbar e}{2mc}\vec{H}$ in Hamiltonian (34) is pseudovector operator which forces on octonic basis of the wave function.

## 7. Nonrelativistic particle in homogeneous magnetic field

Let us find the stationary state of a particle in homogeneous magnetic field. Let the pseudovector of magnetic field intensity is directed along the Z axis:

$$\vec{H} = B\boldsymbol{K}. \tag{37}$$

We select the vector potential in the gauge $(\vec{\nabla},\vec{A}) = 0$:

$$\vec{A} = A_y \boldsymbol{j} = Bx\boldsymbol{j} \ . \tag{38}$$

Then the equation (36) can be written in the following form:

$$-\frac{\hbar^2}{2m}\Delta\breve{\psi} + \frac{\xi\hbar e}{mc}Bx\frac{\partial\breve{\psi}}{\partial y} + \frac{e^2}{2mc^2}B^2x^2\breve{\psi} - \frac{\hbar e}{2mc}B\hat{\boldsymbol{K}}\breve{\psi} = E\breve{\psi} \ . \tag{39}$$

Note that operators $\hat{p}_y = -\xi\hbar\dfrac{\partial}{\partial y}$ and $\hat{p}_z = -\xi\hbar\dfrac{\partial}{\partial z}$ commute with Hamiltonian (34) and all of them have the common eigenfunctions. Therefore we will find the solution of the equation (39) in the following form:

$$\breve{\psi}(\vec{r}) = \breve{W}(x) e^{\frac{\xi}{\hbar}(p_y y + p_z z)} \ , \tag{40}$$

where $p_y$ and $p_z$ are the motion integrals. Substituting (40) into (39), we get

$$\left[-\frac{\hbar^2}{2m}\frac{\partial^2}{\partial x^2} + \frac{p_y^2}{2m} + \frac{p_z^2}{2m} - \frac{ep_y}{mc}Bx + \frac{e^2}{2mc^2}B^2x^2 - \frac{\hbar e}{2mc}B\hat{\boldsymbol{K}}\right]\breve{W} = E\breve{W} \ . \tag{41}$$

Note that Hamiltonian (34) commutes also with octonic pseudovector operator $\hat{\boldsymbol{K}}$, so the solutions of equation (39) are eigenfunctions of the operator $\hat{\boldsymbol{K}}$. In fact it means that in this task there is another quantum number $\lambda$ (eigenvalue of the operator $\hat{\boldsymbol{K}}$) which takes on a value $\lambda = \pm 1$. Therefore we will find the solution of the equation (41) in the form of eigenfunctions of the operator $\hat{\boldsymbol{K}}$ (see (25) and (26)):

$$\breve{W} = F_1^{(\lambda)}(x)(1 + \lambda\boldsymbol{K}) + F_2^{(\lambda)}(x)(\boldsymbol{i} + \lambda\xi\boldsymbol{j}) + F_3^{(\lambda)}(x)(\boldsymbol{E} + \lambda\boldsymbol{k}) + F_4^{(\lambda)}(x)(\boldsymbol{I} + \lambda\xi\boldsymbol{J}) \ . \tag{42}$$

Then the operator in the left part of (41) is scalar, and the equation can be represented in the form

$$-\frac{\hbar^2}{2m}\frac{\partial^2 \breve{W}}{\partial x^2} + \frac{p_y^2}{2m}\breve{W} + \frac{p_z^2}{2m}\breve{W} - \frac{ep_y}{mc}Bx\breve{W} + \frac{e^2}{2mc^2}B^2x^2\breve{W} - \frac{\hbar e}{2mc}B\lambda\breve{W} = E\breve{W} \ . \tag{43}$$

In fact the equation (43) is the system of four identical scalar equations for the functions $F_\alpha^{(\lambda)}(x)$, where $\alpha = 1, 2, 3, 4$ (see (42)). For the fixed $\lambda$ the functions $F_\alpha^{(\lambda)}(x)$ can differ only by a constant.

The equation for each function $F_\alpha^{(\lambda)}(x)$ has the following form:

$$-\frac{\hbar^2}{2m}\frac{\partial^2 F_\alpha^{(\lambda)}}{\partial x^2} + \frac{p_y^2}{2m}F_\alpha^{(\lambda)} + \frac{p_z^2}{2m}F_\alpha^{(\lambda)} - \frac{ep_y}{mc}Bx F_\alpha^{(\lambda)} + \frac{e^2}{2mc^2}B^2x^2 F_\alpha^{(\lambda)} - \frac{\hbar e}{2mc}B\lambda F_\alpha^{(\lambda)} = E F_\alpha^{(\lambda)} \ . \tag{44}$$

After standard algebraic transformations the equation (44) can be represented in the form

$$\frac{\partial^2 F_\alpha^{(\lambda)}}{\partial x^2} + \frac{2m}{\hbar^2}\left\{\left(E - \frac{p_z^2}{2m} + \lambda\frac{\hbar e}{2mc}B\right) - \frac{m}{2}\left(\frac{eB}{mc}\right)^2\left(x - \frac{cp_y}{eB}\right)^2\right\}F_\alpha^{(\lambda)} = 0 \ . \tag{45}$$

On the base of equation (45) we obtain the expression for the energy spectrum of nonrelativistic particle in homogeneous magnetic field:

$$E_{n,\lambda} = \frac{p_z^2}{2m} + \frac{\hbar |e| B}{mc}\left(n + \frac{1}{2}\right) - \lambda\frac{\hbar e}{2mc}B \ . \tag{46}$$

This set of energies is absolutely identical to the energy spectrum obtained from nonrelativistic Pauli equation [26].

Note that if the wave function is the eigenfunction of the operator $\hat{K}$, then some general statements about spatial structure of the wave function can be made. In the stationary state with energy $E$ the wave function has the following form:

$$\breve{\psi}(\vec{r},t) = \breve{\psi}(\vec{r}) e^{-\xi \frac{E}{\hbar} t}. \tag{47}$$

Let the eigenvalue of operator $\hat{K}$ is defined. Then the spatial part of the wave function can be written as the linear combination

$$\breve{\psi}(\vec{r}) = F_1(\vec{r})(1 + \lambda \boldsymbol{K}) + F_2(\vec{r})(\boldsymbol{i} + \lambda \xi \boldsymbol{j}) + F_3(\vec{r})(\boldsymbol{E} + \lambda \boldsymbol{k}) + F_4(\vec{r})(\boldsymbol{I} + \lambda \xi \boldsymbol{J}). \tag{48}$$

The wave function (47) with spatial part (48) has the simple geometrical image. The component $(1 + \lambda \boldsymbol{K})e^{-\frac{\xi}{\hbar} E t}$ is the combination of pseudovector directed parallel to the Z axis and scalar oscillating with frequency $\omega = \frac{E}{\hbar}$. The phase difference between these two oscillations equals to 0 in case of $\lambda = 1$ or $\pi$ in case of $\lambda = -1$. The component $(\boldsymbol{i} + \lambda \xi \boldsymbol{j})e^{-\frac{\xi}{\hbar} E t}$ is the polar vector rotating in the plane perpendicular to the Z axis also with $\omega = \frac{E}{\hbar}$. The direction of the rotating depends on the sign of $\lambda$. When $\lambda = +1$ vector of angular velocity is directed along the Z axis but when $\lambda = -1$ this vector has the opposite direction. The rest components of the wave function (48) have similar interpretation.

## 8. Relations between octonic wave functions and spinors

The spatial structure of the wave functions corresponding to the states with definite eigenvalue $\lambda$ of operator $\hat{K}$ (48) allows to revise the conception of spin and to show relations between spin and space symmetry of the octonic wave function. In this section on the base of octonic wave function we will construct two-component spinor-like objects, which describe spin properties of a particle. It will be shown that spin components of the wave function, generating the basis of two-component spinor, are the special octons with simple spatial structure.

There is some analogy between octonic equation (36) and spinor Pauli equation. As it was shown the energy spectrum (46) is absolutely identical to the spectrum obtained from Pauli equation. However in contrast to Pauli theory the energy levels (46) are defined by eigenvalue $\lambda$ of the octonic spatial operator $\hat{K}$ instead of spin projection onto Z axis.

Besides note that octonic Hamiltonian (34) and Pauli Hamiltonian are coincide except one term. This term in octonic Hamiltonian (34) has the form

$$-\frac{\hbar e}{2mc}\vec{\hat{H}} = -\frac{\hbar e}{2mc}\left(H_x \hat{\boldsymbol{I}} + H_y \hat{\boldsymbol{J}} + H_z \hat{\boldsymbol{K}}\right), \tag{49}$$

and similar term in Pauli Hamiltonian [26] is written by

$$-\frac{\hbar e}{2mc}\left(H_x \hat{\sigma}_x + H_y \hat{\sigma}_y + H_z \hat{\sigma}_z\right), \tag{50}$$

where $\hat{\sigma}_x$, $\hat{\sigma}_y$, $\hat{\sigma}_z$ are Pauli matrices. The comparing (49) and (50) allows to assume the analogy between octonic operators $\hat{\boldsymbol{I}}$, $\hat{\boldsymbol{J}}$, $\hat{\boldsymbol{K}}$ and matrix operators $\hat{\sigma}_x$, $\hat{\sigma}_y$, $\hat{\sigma}_z$. Note that multiplication and commutation rules for these operators are absolutely identical. The difference is that the Pauli matrices force on two-component spinors but $\hat{\boldsymbol{I}}$, $\hat{\boldsymbol{J}}$, $\hat{\boldsymbol{K}}$ operators force on the octonic wave function.

However it can be shown that spinor can be represented as the special case of the octonic wave function and the Pauli matrices realize spinor representation of octonic spatial operators $\hat{I}$, $\hat{J}$, $\hat{K}$.

Let us consider the action of $\hat{I}$ and $\hat{J}$ operators on the simplest eigenfunctions of operator $\hat{K}$ (see (23), (24)):

$$\hat{I}(1+K)=(I-\xi J),\ \hat{I}(I+\xi J)=(1-K),\ \hat{I}(E+k)=(i-\xi j),\ \hat{I}(i+\xi j)=(E-k); \quad (51)$$

$$\hat{I}(1-K)=(I+\xi J),\ \hat{I}(I-\xi J)=(1+K),\ \hat{I}(E-k)=(i+\xi j),\ \hat{I}(i-\xi j)=(E+k); \quad (52)$$

$$\hat{J}(1+K)=\xi(I-\xi J),\ \hat{J}(I+\xi J)=\xi(1-K),\ \hat{J}(E+k)=\xi(i-\xi j),\ \hat{J}(i+\xi j)=\xi(E-k); \quad (53)$$

$$\hat{J}(1-K)=-\xi(I+\xi J),\ \hat{J}(I-\xi J)=-\xi(1+K),\ \hat{J}(E-k)=-\xi(i+\xi j),\ \hat{J}(i-\xi j)=-\xi(E+k). \quad (54)$$

It is seen from (51)-(54) that the simplest eigenfunctions of the operator $\hat{K}$ generate the closed system relatively octonic operators $\hat{I}$, $\hat{J}$, $\hat{K}$. Note that the components of spinors are transformed each through other in a similar way by means of Pauli operators.

Here we will show how the octonic basis of spinor can be constructed from the simplest eigenfunctions of the operator $\hat{K}$. Let us require that the action of $\hat{I}$, $\hat{J}$ and $\hat{K}$ on basis functions of spinor in octonic representation is equivalent to action of matrix operators

$$\hat{\sigma}_x = \begin{pmatrix} 0 & 1 \\ 1 & 0 \end{pmatrix}, \qquad \hat{\sigma}_y = \begin{pmatrix} 0 & -\xi \\ \xi & 0 \end{pmatrix}, \qquad \hat{\sigma}_z = \begin{pmatrix} 1 & 0 \\ 0 & -1 \end{pmatrix} \quad (55)$$

on the basis function of spinor in spinor representation. Then matrix operators $\hat{\sigma}_x$, $\hat{\sigma}_y$, $\hat{\sigma}_z$ will realize the spinor representation of spatial operators $\hat{I}$, $\hat{J}$, $\hat{K}$.

On the base of advanced assumptions the basis functions of spinor can be written as the linear combination of simplest functions of the operator $\hat{K}$.

$$\begin{pmatrix} 1 \\ 0 \end{pmatrix} = \alpha_1(1+K) + \alpha_2(i+\xi j) + \alpha_3(E+k) + \alpha_4(I+\xi J), \quad (56)$$

$$\begin{pmatrix} 0 \\ 1 \end{pmatrix} = \beta_1(1-K) + \beta_2(i-\xi j) + \beta_3(E-k) + \beta_4(I-\xi J). \quad (57)$$

Let force on (56) and (57) by operators $\hat{\sigma}_x = \hat{I}$ and $\hat{\sigma}_y = \hat{J}$. At that taking into account (56)-(57) we obtain the following octonic equations in spinor form

$$\begin{cases} \begin{pmatrix} 0 & 1 \\ 1 & 0 \end{pmatrix}\begin{pmatrix} 1 \\ 0 \end{pmatrix} = \begin{pmatrix} 0 \\ 1 \end{pmatrix}, \\ \begin{pmatrix} 0 & 1 \\ 1 & 0 \end{pmatrix}\begin{pmatrix} 0 \\ 1 \end{pmatrix} = \begin{pmatrix} 1 \\ 0 \end{pmatrix}, \\ \begin{pmatrix} 0 & -\xi \\ \xi & 0 \end{pmatrix}\begin{pmatrix} 1 \\ 0 \end{pmatrix} = \xi\begin{pmatrix} 0 \\ 1 \end{pmatrix}, \\ \begin{pmatrix} 0 & -\xi \\ \xi & 0 \end{pmatrix}\begin{pmatrix} 0 \\ 1 \end{pmatrix} = -\xi\begin{pmatrix} 1 \\ 0 \end{pmatrix}. \end{cases} \quad (58)$$

Then the system of matrix equalities (58) in octonic representation leads to the equations for arbitrary coefficients in (56) and (57):

$$\begin{cases} \hat{I}\{\alpha_1(1+K)+\alpha_2(i+\xi j)+\alpha_3(E+k)+\alpha_4(I+\xi J)\} = \beta_1(1-K)+\beta_2(i-\xi j)+\beta_3(E-k)+\beta_4(I-\xi J), \\ \hat{I}\{\beta_1(1-K)+\beta_2(i-\xi j)+\beta_3(E-k)+\beta_4(I-\xi J)\} = \alpha_1(1+K)+\alpha_2(i+\xi j)+\alpha_3(E+k)+\alpha_4(I+\xi J), \\ \hat{J}\{\alpha_1(1+K)+\alpha_2(i+\xi j)+\alpha_3(E+k)+\alpha_4(I+\xi J)\} = \xi\{\beta_1(1-K)+\beta_2(i-\xi j)+\beta_3(E-k)+\beta_4(I-\xi J)\}, \\ \hat{J}\{\beta_1(1-K)+\beta_2(i-\xi j)+\beta_3(E-k)+\beta_4(I-\xi J)\} = -\xi\{\alpha_1(1+K)+\alpha_2(i+\xi j)+\alpha_3(E+k)+\alpha_4(I+\xi J)\}. \end{cases} \quad (59)$$

Taking into account (51)-(54) we get the following system of equations:

$$\begin{cases} \alpha_1(I-\xi J)+\alpha_2(E-k)+\alpha_3(i-\xi j)+\alpha_4(1-K) = \beta_1(1-K)+\beta_2(i-\xi j)+\beta_3(E-k)+\beta_4(I-\xi J), \\ \beta_1(I+\xi J)+\beta_2(E+k)+\beta_3(i+\xi j)+\beta_4(1+K) = \alpha_1(1+K)+\alpha_2(i+\xi j)+\alpha_3(E+k)+\alpha_4(I+\xi J), \\ \alpha_1\xi(I-\xi J)+\alpha_2\xi(E-k)+\alpha_3\xi(i-\xi j)+\alpha_4\xi(1-K) = \beta_1\xi(1-K)+\beta_2\xi(i-\xi j)+\beta_3\xi(E-k)+\beta_4\xi(I-\xi J), \\ -\beta_1\xi(I+\xi J)-\beta_2\xi(E+k)-\beta_3\xi(i+\xi j)-\beta_4\xi(1+K) = -\alpha_1\xi(1+K)-\alpha_2\xi(i+\xi j)-\alpha_3\xi(E+k)-\alpha_4\xi(I+\xi J). \end{cases} \quad (60)$$

The system (60) is equivalent to the system of four scalar equations

$$\begin{cases} \beta_4 = \alpha_1, \\ \beta_3 = \alpha_2, \\ \beta_2 = \alpha_3, \\ \beta_1 = \alpha_4, \end{cases} \quad (61)$$

which define the conditions when eigenfunctions (56), (57) are transformed under $\hat{I}$, $\hat{J}$, $\hat{K}$ operators like a spinor.

Then for any complex values $\alpha_1$, $\alpha_2$, $\alpha_3$, $\alpha_4$ the functions

$$\begin{pmatrix} 1 \\ 0 \end{pmatrix} = \alpha_1(1+K)+\alpha_2(i+\xi j)+\alpha_3(E+k)+\alpha_4(I+\xi J), \quad (62)$$

$$\begin{pmatrix} 0 \\ 1 \end{pmatrix} = \alpha_4(1-K)+\alpha_3(i-\xi j)+\alpha_2(E-k)+\alpha_1(I-\xi J) \quad (63)$$

can be used as the octonic basis functions of spinor.

However as it is seen from (62) and (63) this definition of spinor basis functions contains the ambiguity connected with arbitrary coefficients $\alpha_1$, $\alpha_2$, $\alpha_3$, $\alpha_4$. It means that the octonic wave function includes more information about particle state than the wave function in the spinor representation. In fact the spinor function includes information only about spin projection.

Thus the application of spatial operators and octonic wave function allows to describe the spin properties of a particle and to establish correspondence between spinor and octonic representations. Moreover the application of octon's algebra allows to go out of the frames of the Pauli theory and to discover unusual space-time structure of the wave function. In octonic quantum mechanics the wave function of the particle with definite spin projection is the octonic oscillator (47)-(48) with longitudinal linear and transversal circular spatial polarizations (see section 6).

## 9. Relativistic particle in homogeneous magnetic field

Let consider the relativistic particle in an external electromagnetic field directed along the Z axis: $\vec{H}=B\boldsymbol{K}$, $\vec{A}=A_y\boldsymbol{j}=Bx\boldsymbol{j}$ and assume the condition $(\vec{\nabla},\vec{A})=0$. Then octonic equation for relativistic particle (30) can be written as

$$\frac{1}{c^2}\frac{\partial^2 \breve{\psi}}{\partial t^2}-\Delta\breve{\psi}+\frac{2\xi e}{\hbar c}Bx\frac{\partial \breve{\psi}}{\partial y}+\frac{m^2c^2}{\hbar^2}\breve{\psi}+\frac{e^2}{\hbar^2 c^2}B^2x^2\breve{\psi}-\frac{e}{\hbar c}\vec{H}\,\breve{\psi}=0. \quad (64)$$

For the stationary state with the energy $E$ we get

$$\left[ -\Delta + \frac{2\xi e}{\hbar c} Bx \frac{\partial}{\partial y} + \frac{m^2 c^2}{\hbar^2} + \frac{e^2}{\hbar^2 c^2} B^2 x^2 - \frac{e}{\hbar c} B\hat{K} \right] \breve{\psi} = \frac{E^2}{\hbar^2 c^2} \breve{\psi}. \tag{65}$$

This equation can be considered as the equation on the eigenvalues and eigenfunctions of complicated operator placed in the left part. Since this operator commutes with operators $\hat{p}_y = -\xi\hbar \frac{\partial}{\partial y}$ and $\hat{p}_z = -\xi\hbar \frac{\partial}{\partial z}$ all of them have the general system of eigenfunctions. Therefore we will find the solution of (65) in the form

$$\breve{\psi} = \breve{W}(x) e^{\frac{\xi}{\hbar}(p_y y + p_z z)}, \tag{66}$$

where $p_y$ and $p_z$ are the motion integrals. Substituting (66) into (65) we get

$$\left[ \frac{p_y^2}{\hbar^2} + \frac{p_z^2}{\hbar^2} - \frac{\partial^2}{\partial x^2} - \frac{2e p_y}{\hbar^2 c} Bx + \frac{m^2 c^2}{\hbar^2} + \frac{e^2}{\hbar^2 c^2} B^2 x^2 - \frac{e}{\hbar c} B\hat{K} \right] \breve{W} = \frac{E^2}{\hbar^2 c^2} \breve{W}. \tag{67}$$

Note that operator in the left part of (67) commutes also with $\hat{K}$, so we can find the solution as the eigenfunctions of the operator $\hat{K}$ (25)-(26):

$$\breve{W} = F_1^{(\lambda)}(x)(1 + \lambda \mathbf{K}) + F_2^{(\lambda)}(x)(\mathbf{i} + \lambda \xi \mathbf{j}) + F_3^{(\lambda)}(x)(\mathbf{E} + \lambda \mathbf{k}) + F_4^{(\lambda)}(x)(\mathbf{I} + \lambda \xi \mathbf{J}). \tag{68}$$

So we get scalar equations for the functions $F_\alpha^{(\lambda)}(x)$ (where $\alpha = 1, 2, 3, 4$):

$$\frac{p_y^2}{\hbar^2} F_\alpha^{(\lambda)} + \frac{p_z^2}{\hbar^2} F_\alpha^{(\lambda)} - \frac{\partial^2 F_\alpha^{(\lambda)}}{\partial x^2} - \frac{2e p_y}{\hbar^2 c} Bx F_\alpha^{(\lambda)} + \frac{m^2 c^2}{\hbar^2} F_\alpha^{(\lambda)} + \frac{e^2}{\hbar^2 c^2} B^2 x^2 F_\alpha^{(\lambda)} - \frac{e}{\hbar c} B\lambda F_\alpha^{(\lambda)} = \frac{E^2}{\hbar^2 c^2} F_\alpha^{(\lambda)}. \tag{69}$$

After some algebraic transformations the equation (69) can be represented as

$$\frac{\partial^2 F_\alpha^{(\lambda)}}{\partial x^2} + \left[ \left( \frac{E^2}{\hbar^2 c^2} - \frac{p_z^2}{\hbar^2} - \frac{m^2 c^2}{\hbar^2} + \lambda \frac{eB}{\hbar c} \right) - \left( \frac{eB}{\hbar c} \right)^2 \left( x - \frac{c p_y}{eB} \right)^2 \right] F_\alpha^{(\lambda)} = 0. \tag{70}$$

Then the energy spectrum is defined by the following expression:

$$E_{n,\lambda}^2 = m^2 c^4 + p_z^2 c^2 + |e| B\hbar c (2n+1) - \lambda e B\hbar c. \tag{71}$$

Note that in the nonrelativistic limit the energy levels (71) are transformed in (46) for the levels counted out of rest energy. The spatial structure of the wave functions is defined by analogy with nonrelativistic case (see (47)-(48)) and also corresponds to octonic oscillator. Emphasize that the octonic wave function of the relativistic particle can be represented also in the form of two component spinor as for nonrelativistic particle.

## 10. Relations between octonic wave functions and bispinors

In the Dirac theory spin operators are represented by four-row matrices and wave functions by four-component bispinor. Two components of bispinor describe a particle and other two components describe an antiparticle. The second order equation in the Dirac theory is written in the following form [25, 27]:

$$\left[ -\Delta + \frac{1}{c^2} \frac{\partial^2}{\partial t^2} + \frac{2\xi e}{\hbar c} \left( (\vec{A}, \vec{\nabla}) + \frac{\Phi}{c} \frac{\partial}{\partial t} \right) + \frac{m^2 c^2}{\hbar^2} + \frac{e^2}{\hbar^2 c^2} (A^2 - \Phi^2) \right] \psi - \frac{e}{\hbar c} \hat{\vec{\Sigma}} \vec{H} \, \psi + \frac{\xi e}{\hbar c} \hat{\vec{\alpha}} \vec{E} \, \psi = 0. \tag{72}$$

Here the components of the matrix vectors $\hat{\vec{\Sigma}}$ and $\hat{\vec{\alpha}}$ have the following form:

$$\hat{\Sigma}_x = \begin{pmatrix} 0 & 1 & 0 & 0 \\ 1 & 0 & 0 & 0 \\ 0 & 0 & 0 & 1 \\ 0 & 0 & 1 & 0 \end{pmatrix}, \qquad \hat{\Sigma}_y = \begin{pmatrix} 0 & -\xi & 0 & 0 \\ \xi & 0 & 0 & 0 \\ 0 & 0 & 0 & -\xi \\ 0 & 0 & \xi & 0 \end{pmatrix}, \qquad \hat{\Sigma}_z = \begin{pmatrix} 1 & 0 & 0 & 0 \\ 0 & -1 & 0 & 0 \\ 0 & 0 & 1 & 0 \\ 0 & 0 & 0 & -1 \end{pmatrix}; \qquad (73)$$

$$\hat{\alpha}_x = \begin{pmatrix} 0 & 1 & 0 & 0 \\ 1 & 0 & 0 & 0 \\ 0 & 0 & 0 & -1 \\ 0 & 0 & -1 & 0 \end{pmatrix}, \qquad \hat{\alpha}_y = \begin{pmatrix} 0 & -\xi & 0 & 0 \\ \xi & 0 & 0 & 0 \\ 0 & 0 & 0 & \xi \\ 0 & 0 & -\xi & 0 \end{pmatrix}, \qquad \hat{\alpha}_z = \begin{pmatrix} 1 & 0 & 0 & 0 \\ 0 & -1 & 0 & 0 \\ 0 & 0 & -1 & 0 \\ 0 & 0 & 0 & 1 \end{pmatrix}. \qquad (74)$$

Comparing equations (72) and (30) one can assume that matrices (73) and (74) realize bispinor representation of octonic spatial operators:

$$\hat{\Sigma}_x = \hat{I}, \qquad \hat{\Sigma}_y = \hat{J}, \qquad \hat{\Sigma}_z = \hat{K}, \qquad \hat{\alpha}_x = \hat{i}, \qquad \hat{\alpha}_y = \hat{j}, \qquad \hat{\alpha}_z = \hat{k}. \qquad (75)$$

Note that commutation and multiplication rules for matrix operators (73)-(74) are absolutely identical to the rules for octonic spatial operators.

By analogy with spinor representation of the octonic wave function developed in section 6 it can be shown that the bispinor wave function of the Dirac theory also can be represented as the particular case of octonic wave function. Let us construct the octonic representation of basis functions of bispinor as it was done previously for spinors.

First we note that the operator $\hat{\Sigma}_z$ is diagonal in the bispinor representation. It means (see (75)) that the octonic basis functions of bispinor are the eigenfunctions of the operator $\hat{K}$. Then basis functions of bispinor (we will indicate them for short as $\chi_s$, $s = 1, 2, 3, 4$) can be represented in the form of linear combination of the simplest function of the operator $\hat{K}$:

$$\chi_1 = \begin{pmatrix} 1 \\ 0 \\ 0 \\ 0 \end{pmatrix} = a_1(1+K) + a_2(I+\xi J) + a_3(E+k) + a_4(i+\xi j),$$

$$\chi_2 = \begin{pmatrix} 0 \\ 1 \\ 0 \\ 0 \end{pmatrix} = b_1(1-K) + b_2(I-\xi J) + b_3(E-k) + b_4(i-\xi j),$$

$$\chi_3 = \begin{pmatrix} 0 \\ 0 \\ 1 \\ 0 \end{pmatrix} = A_1(1+K) + A_2(I+\xi J) + A_3(E+k) + A_4(i+\xi j), \qquad (76)$$

$$\chi_4 = \begin{pmatrix} 0 \\ 0 \\ 0 \\ 1 \end{pmatrix} = B_1(1-K) + B_2(I-\xi J) + B_3(E-k) + B_4(i-\xi j),$$

where $a_s$, $b_s$, $A_s$, $B_s$ ($s = 1, 2, 3, 4$) are arbitrary complex coefficients.

The requirement that operators $\hat{I}$, $\hat{J}$, $\hat{K}$ act on the octonic wave function identically as operators $\hat{\Sigma}_x$, $\hat{\Sigma}_y$, $\hat{\Sigma}_z$ act on the corresponding bispinor is satisfied only in the case if coefficients $a_s$, $b_s$, $A_s$, $B_s$ are satisfied the following conditions:

$$\begin{cases} a_1 = b_2, a_2 = b_1, a_3 = b_4, a_4 = b_3, \\ A_1 = B_2, A_2 = B_1, A_3 = B_4, A_4 = B_3. \end{cases} \quad (77)$$

Analogous requirements for the operators $\hat{i}$, $\hat{j}$, $\hat{k}$ and matrices $\hat{\alpha}_x$, $\hat{\alpha}_y$, $\hat{\alpha}_z$ lead to the conditions

$$\begin{cases} a_1 = a_3, a_2 = a_4, \\ A_1 = -A_3, A_2 = -A_4. \end{cases} \quad (78)$$

As a result the basis functions of bispinor have the following structure:

$$\chi_1 = a(1 + K + E + k) + b(I + \xi J + i + \xi j),$$

$$\chi_2 = b(1 - K + E - k) + a(I - \xi J + i - \xi j),$$

$$\chi_3 = A(1 + K - E - k) + B(I + \xi J - i - \xi j), \quad (79)$$

$$\chi_4 = B(1 - K - E + k) + A(I - \xi J - i + \xi j).$$

Here we used symbols $a = a_1$, $b = a_2$, $A = A_3$, $B = A_4$.

Except matrices $\hat{\alpha}_x$, $\hat{\alpha}_y$, $\hat{\alpha}_z$ in the Dirac theory they consider the matrix $\hat{\beta}$, which has the following form:

$$\hat{\beta} = \begin{pmatrix} 0 & 0 & 1 & 0 \\ 0 & 0 & 0 & 1 \\ 1 & 0 & 0 & 0 \\ 0 & 1 & 0 & 0 \end{pmatrix}. \quad (80)$$

Let us find an octonic operator, for which the bispinor representation is realized by the matrix $\hat{\beta}$. Note that the rules of commutation for operator of spatial inversion $\hat{R}$ and operators $\hat{i}$, $\hat{j}$, $\hat{k}$, $\hat{I}$, $\hat{J}$, $\hat{K}$ coincide with commutation rules between matrices $\hat{\beta}$ and $\hat{\alpha}_x$, $\hat{\alpha}_y$, $\hat{\alpha}_z$, $\hat{\Sigma}_x$, $\hat{\Sigma}_y$, $\hat{\Sigma}_z$. Let us assume that $\hat{R} = \hat{\beta}$. This imposes the additional conditions on the coefficients in basis functions of bispinor:

$$\begin{cases} a = A, \\ b = B. \end{cases} \quad (81)$$

Taking into account (81) for the structure of basis functions we get

$$\chi_1 = a(1 + K + E + k) + b(I + \xi J + i + \xi j),$$

$$\chi_2 = b(1 - K + E - k) + a(I - \xi J + i - \xi j),$$

$$\chi_3 = a(1 + K - E - k) + b(I + \xi J - i - \xi j), \quad (82)$$

$$\chi_4 = b(1 - K - E + k) + a(I - \xi J - i + \xi j).$$

Thus in the frame of made assumptions bispinor is the column of coefficients of octonic wave function expansion on the basis (82):

$$\begin{pmatrix} W_1 \\ W_2 \\ W_3 \\ W_4 \end{pmatrix} = W_1 \chi_1 + W_2 \chi_2 + W_3 \chi_3 + W_4 \chi_4. \qquad (83)$$

Using the basis functions (82) one can examine that the operator $\hat{E}$ in the bispinor representation has the following matrix form:

$$\hat{E} = \begin{pmatrix} 1 & 0 & 0 & 0 \\ 0 & 1 & 0 & 0 \\ 0 & 0 & -1 & 0 \\ 0 & 0 & 0 & -1 \end{pmatrix}. \qquad (84)$$

It is clearly seen from (84) that the states corresponding to the particle and antiparticle in the Dirac theory are the eigenstates of the operator $\hat{E}$ in the frame of octonic quantum mechanics. Thus we have shown that the rules of correspondence (75) between octonic spatial operators and matrix operators $\hat{\alpha}_x$, $\hat{\alpha}_y$, $\hat{\alpha}_z$, $\hat{\Sigma}_x$, $\hat{\Sigma}_y$, $\hat{\Sigma}_z$ realize the bispinor representation of the octonic wave function.

Then we can set accordance between Dirac matrices $\hat{\gamma}_i$ ($i = 0,1,2,3$) defined as [25]

$$\hat{\gamma}_0 = \hat{\beta}, \qquad \hat{\gamma}_i = \hat{\gamma}_0 \hat{\alpha}_i, \qquad (85)$$

and the following octonic spatial operators:

$$\hat{\gamma}_0 = \hat{R}, \qquad \hat{\gamma}_1 = \hat{R}\hat{\boldsymbol{i}}, \qquad \hat{\gamma}_2 = \hat{R}\hat{\boldsymbol{j}}, \qquad \hat{\gamma}_3 = \hat{R}\hat{\boldsymbol{k}}. \qquad (86)$$

It is easy to check that the matrix $\hat{\gamma}_5$ defined as

$$\hat{\gamma}_5 = \xi \hat{\gamma}_0 \hat{\gamma}_1 \hat{\gamma}_2 \hat{\gamma}_3 = -\xi \hat{\alpha}_x \hat{\alpha}_y \hat{\alpha}_z, \qquad (87)$$

coincides with the matrix of pseudoscalar operator $\hat{E}$ in the bispinor representation (84):

$$\hat{\gamma}_5 = \hat{E}. \qquad (88)$$

Note that the definition of the matrix $\hat{\gamma}_5$ (87) corresponds to the definition of $E$ via unit vectors of octon's basis (see [24]).

The basis functions (82) of bispinor can be represented also in the following form:

$$\chi_1 = \frac{(1+E)}{2} \left\{ f \frac{(1+K)}{2} + g \frac{(I+\xi J)}{2} \right\},$$

$$\chi_2 = \frac{(1+E)}{2} \left\{ g \frac{(1-K)}{2} + f \frac{(I-\xi J)}{2} \right\},$$

$$\chi_3 = \frac{(1-E)}{2} \left\{ f \frac{(1+K)}{2} + g \frac{(I+\xi J)}{2} \right\}, \qquad (89)$$

$$\chi_4 = \frac{(1-E)}{2} \left\{ g \frac{(1-K)}{2} + f \frac{(I-\xi J)}{2} \right\}.$$

The coefficients $\frac{1}{2}$ are put in the basis functions (89) for convenience simultaneously with redefinition of arbitrary constants: $f = 4a$, $g = 4b$. The new constants $f$ and $g$ can be considered as some quantum numbers.

The octonic operators

$$\hat{a}_+ = \frac{1+\hat{E}}{2}, \quad \hat{a}_- = \frac{1-\hat{E}}{2}, \tag{90}$$

with following properties

$$\hat{a}_+^2 = \hat{a}_+, \quad \hat{a}_-^2 = \hat{a}_-, \quad \hat{a}_+\hat{a}_- = \hat{a}_-\hat{a}_+ = 0, \quad \hat{a}_+ + \hat{a}_- = 1, \tag{91}$$

are the octonic projection operators. Operator $\hat{a}_+$ extracts from bispinor the states corresponding to the particle and operator $\hat{a}_-$ extracts states corresponding to the antiparticle:

$$\hat{a}_+ \begin{pmatrix} W_1 \\ W_2 \\ W_3 \\ W_4 \end{pmatrix} = \begin{pmatrix} W_1 \\ W_2 \\ 0 \\ 0 \end{pmatrix}, \quad \hat{a}_- \begin{pmatrix} W_1 \\ W_2 \\ W_3 \\ W_4 \end{pmatrix} = \begin{pmatrix} 0 \\ 0 \\ W_3 \\ W_4 \end{pmatrix}. \tag{92}$$

The projection operators $\hat{a}_+$ and $\hat{a}_-$ allow to extract the components corresponding to the particle and antiparticle from any wave function. Let us consider the arbitrary octonic wave function in the form $\breve{\psi} = \psi_0 + \vec{\psi} + \tilde{\varphi}_0 + \vec{\tilde{\varphi}}$. Then we get

$$\hat{a}_+\breve{\psi} = \hat{a}_+\left(\psi_0 + \vec{\psi} + \tilde{\varphi}_0 + \vec{\tilde{\varphi}}\right) = \hat{a}_+\left(\psi_0 + \vec{\psi} + \varphi_0 \boldsymbol{E} + \vec{\varphi}\boldsymbol{E}\right) = \left(\frac{1+\boldsymbol{E}}{2}\right)\left(\psi_0 + \varphi_0 + \vec{\psi} + \vec{\varphi}\right), \tag{93}$$

$$\hat{a}_-\breve{\psi} = \hat{a}_-\left(\psi_0 + \vec{\psi} + \tilde{\varphi}_0 + \vec{\tilde{\varphi}}\right) = \hat{a}_-\left(\psi_0 + \vec{\psi} + \varphi_0 \boldsymbol{E} + \vec{\varphi}\boldsymbol{E}\right) = \left(\frac{1-\boldsymbol{E}}{2}\right)\left(\psi_0 + \varphi_0 + \vec{\psi} + \vec{\varphi}\right). \tag{94}$$

Here we write in an explicit form the scalar $\varphi_0$ in the pseudoscalar part $\tilde{\varphi}_0 = \boldsymbol{E}\varphi_0$ and vector $\vec{\varphi}$ in the pseudovector part $\vec{\tilde{\varphi}} = \boldsymbol{E}\vec{\varphi}$ of the wave function. On the base of expressions (93) and (94) we can write the general form of octonic wave function for particle $\breve{\psi}_p$ and for antiparticle $\breve{\psi}_a$. The scalar part of $\breve{\psi}_p$ equals to the pseudoscalar part and vector part equals to the pseudovector part. On the other hand the scalar part of $\breve{\psi}_a$ equals to the pseudoscalar part with the opposite sign, and vector part equals to the pseudovector part with the opposite sign too:

$$\breve{\psi}_p = \left(\frac{1+\boldsymbol{E}}{2}\right)\left(\psi_{0p} + \vec{\psi}_p\right), \tag{95}$$

$$\breve{\psi}_a = \left(\frac{1-\boldsymbol{E}}{2}\right)\left(\psi_{0a} + \vec{\psi}_a\right). \tag{96}$$

Here index $p$ corresponds to the particle and index $a$ corresponds to the antiparticle.

By analogy with projection operators (90) one can construct octonic operators

$$\hat{b}_+ = \frac{1+\hat{R}}{2}, \quad \hat{b}_- = \frac{1-\hat{R}}{2}, \tag{97}$$

which extract from the wave function the symmetric and antisymmetric parts relatively of spatial inversion:

$$\hat{b}_+(\psi_0 + \vec{\psi} + \tilde{\varphi}_0 + \vec{\varphi}) = \psi_0 + \vec{\varphi},$$
$$\hat{b}_-(\psi_0 + \vec{\psi} + \tilde{\varphi}_0 + \vec{\varphi}) = \tilde{\varphi}_0 + \vec{\psi}.$$
(98)

Moreover the octonic projection operators

$$\hat{s}_+ = \frac{1+\hat{K}}{2}, \qquad \hat{s}_- = \frac{1-\hat{K}}{2} \qquad (99)$$

extract the states with defined spin projection on the Z axis from the octonic wave function.

The octonic basis of bispinor (89) is written in representation where operators $\hat{K}$ and $\hat{E}$ are diagonal. However one can construct the representation where two other commuting operators will be diagonal. For example let us construct the representation where operators $\hat{K}$ and $\hat{R}$ will have the diagonal form (so-called standard representation [25]). In this representation the octonic operators have the following matrix form:

$$\hat{I} = \begin{pmatrix} 0 & 1 & 0 & 0 \\ 1 & 0 & 0 & 0 \\ 0 & 0 & 0 & 1 \\ 0 & 0 & 1 & 0 \end{pmatrix}, \qquad \hat{J} = \begin{pmatrix} 0 & -\xi & 0 & 0 \\ \xi & 0 & 0 & 0 \\ 0 & 0 & 0 & -\xi \\ 0 & 0 & \xi & 0 \end{pmatrix}, \qquad \hat{K} = \begin{pmatrix} 1 & 0 & 0 & 0 \\ 0 & -1 & 0 & 0 \\ 0 & 0 & 1 & 0 \\ 0 & 0 & 0 & -1 \end{pmatrix}; \qquad (100)$$

$$\hat{i} = \begin{pmatrix} 0 & 0 & 0 & 1 \\ 0 & 0 & 1 & 0 \\ 0 & 1 & 0 & 0 \\ 1 & 0 & 0 & 0 \end{pmatrix}, \qquad \hat{j} = \begin{pmatrix} 0 & 0 & 0 & -\xi \\ 0 & 0 & \xi & 0 \\ 0 & -\xi & 0 & 0 \\ \xi & 0 & 0 & 0 \end{pmatrix}, \qquad \hat{k} = \begin{pmatrix} 0 & 0 & 1 & 0 \\ 0 & 0 & 0 & -1 \\ 1 & 0 & 0 & 0 \\ 0 & -1 & 0 & 0 \end{pmatrix}, \qquad (101)$$

$$\hat{E} = \begin{pmatrix} 0 & 0 & 1 & 0 \\ 0 & 0 & 0 & 1 \\ 1 & 0 & 0 & 0 \\ 0 & 1 & 0 & 0 \end{pmatrix}, \qquad \hat{R} = \begin{pmatrix} 1 & 0 & 0 & 0 \\ 0 & 1 & 0 & 0 \\ 0 & 0 & -1 & 0 \\ 0 & 0 & 0 & -1 \end{pmatrix}. \qquad (102)$$

Reasoning by analogy with (76)-(89) we get the following expressions for the octonic basis functions of bispinor in standard representation:

$$\chi_1 = \left\{ f' \frac{(1+K)}{2} + g' \frac{(I+\xi J)}{2} \right\},$$
$$\chi_2 = \left\{ g' \frac{(1-K)}{2} + f' \frac{(I-\xi J)}{2} \right\},$$
$$\chi_3 = E \left\{ f' \frac{(1+K)}{2} + g' \frac{(I+\xi J)}{2} \right\},$$
$$\chi_4 = E \left\{ f' \frac{(1+K)}{2} + g' \frac{(I+\xi J)}{2} \right\}.$$
(103)

Thus in the frames of octonic quantum mechanics it was managed to set correspondence between bispinor and octonic description of particles. It was shown that the information about the states corresponding to the particle and antiparticle in the Dirac theory is contained directly in the spatial structure of the wave function. These states have different symmetry relatively permutation of scalar and pseudoscalar as well as vector and pseudovector components of the wave function: the symmetric function corresponds to the particle but antisymmetric function corresponds to the antiparticle. Besides the states of particle and antiparticle correspond to the different eigenvalues of spatial pseudoscalar operator $\hat{E}$. Projection operators, which extract the states of particle and antiparticle from the wave function, are also spatial octonic operators.

In fact from the position of octonic quantum mechanics in the Pauli spinor representation they separate two parts of the wave function which correspond to different spin projections on the definite direction. In the Dirac theory they do the additional separation of the wave function corresponding to the states of particle and antiparticle.

However the application of octon's algebra allows to go out of the frames of the bispinor Dirac representation and to discover additional internal degree of freedom in the wave function. From expressions (89) and (103) it is clearly seen that the wave function has another internal parameter, which also has spatial nature and characterizes the amplitude relation between different polarizations of the octonic oscillator.

## 11. Operators of spatial reflections and discrete turns

In the section 3 we considered spatial operators $\hat{i}$, $\hat{j}$, $\hat{k}$, $\hat{E}$, $\hat{I}$, $\hat{J}$, $\hat{K}$, which in fact permute the components of the wave function. However these operators do not exhaust all discrete spatial transformation of the wave function. Here we will consider in addition the operators of discrete turns on angle $\pi$ around axes of Cartesian coordinate system and operators of spatial reflections in the coordinate planes. The action of these operators leads only to changing the coordinate axes directions and consequently changes signs at different components of the wave function.

Let us indicate the operators of reflection in the planes perpendicular to the $X$, $Y$, $Z$ axes as $\hat{R}_x$, $\hat{R}_y$, $\hat{R}_z$ and corresponding turn operators as $\hat{\pi}_x$, $\hat{\pi}_y$, $\hat{\pi}_z$. These operators generate closed commutative algebra of turns and reflections, which is represented for information in appendix 4. Note that the operator of spatial inversion $\hat{R}$ is the element of this algebra and can be represented in the following form:

$$\hat{R} = \hat{R}_x \hat{R}_y \hat{R}_z. \tag{104}$$

The rules of action of operators $\hat{\pi}_x$, $\hat{\pi}_y$, $\hat{\pi}_z$, $\hat{R}_x$, $\hat{R}_y$, $\hat{R}_z$ on the elements of octon's basis are represented in the table 2.

*Table 2. The action of operators of reflections and discrete turns on the elements of octon's basis.*

|             | $i$  | $j$  | $k$  | $E$  | $I$  | $J$  | $K$  |
|-------------|------|------|------|------|------|------|------|
| $\hat{R}_x$ | $-i$ | $j$  | $k$  | $-E$ | $I$  | $-J$ | $-K$ |
| $\hat{R}_y$ | $i$  | $-j$ | $k$  | $-E$ | $-I$ | $J$  | $-K$ |
| $\hat{R}_z$ | $i$  | $j$  | $-k$ | $-E$ | $-I$ | $-J$ | $K$  |
| $\hat{R}$   | $-i$ | $-j$ | $-k$ | $-E$ | $I$  | $J$  | $K$  |
| $\hat{\pi}_x$ | $i$  | $-j$ | $-k$ | $E$  | $I$  | $-J$ | $-K$ |
| $\hat{\pi}_y$ | $-i$ | $j$  | $-k$ | $E$  | $-I$ | $J$  | $-K$ |
| $\hat{\pi}_z$ | $-i$ | $-j$ | $k$  | $E$  | $-I$ | $-J$ | $K$  |

For example, the reflection in the plane YZ is described by the $\hat{R}_x$ operator:

$$\hat{R}_x \left( \psi_0 + \psi_1 \boldsymbol{i} + \psi_2 \boldsymbol{j} + \psi_3 \boldsymbol{k} + \varphi_0 \boldsymbol{E} + \varphi_1 \boldsymbol{I} + \varphi_2 \boldsymbol{J} + \varphi_3 \boldsymbol{K} \right) = \\ = \psi_0 - \psi_1 \boldsymbol{i} + \psi_2 \boldsymbol{j} + \psi_3 \boldsymbol{k} - \varphi_0 \boldsymbol{E} + \varphi_1 \boldsymbol{I} - \varphi_2 \boldsymbol{J} - \varphi_3 \boldsymbol{K} \ . \quad (105)$$

The commutation rules for $\hat{\pi}_x$, $\hat{\pi}_y$, $\hat{\pi}_z$, $\hat{R}_x$, $\hat{R}_y$, $\hat{R}_z$ and $\hat{\boldsymbol{i}}$, $\hat{\boldsymbol{j}}$, $\hat{\boldsymbol{k}}$, $\hat{\boldsymbol{E}}$, $\hat{\boldsymbol{I}}$, $\hat{\boldsymbol{J}}$, $\hat{\boldsymbol{K}}$ operators are represented in appendix 5.

Let us consider the eigenvalues and eigenfunctions of the $\hat{\pi}_z$ operator. The eigenvalues $\lambda$ of $\hat{\pi}_z$ are $\lambda = \pm 1$ and corresponding primary eigenfunctions have the following form:

$$\lambda = +1: \qquad 1,\ \boldsymbol{E},\ \boldsymbol{k},\ \boldsymbol{K}\ ; \quad (106)$$

$$\lambda = -1: \qquad \boldsymbol{i},\ \boldsymbol{j},\ \boldsymbol{I},\ \boldsymbol{J}\ . \quad (107)$$

Note that the eigenfunctions can be chosen also as the liner combinations of corresponding functions (106)-(107):

$$\lambda = +1: \qquad (1+\boldsymbol{K}),\ (1-\boldsymbol{K}),\ (\boldsymbol{E}+\boldsymbol{k}),\ (\boldsymbol{E}-\boldsymbol{k})\ ; \quad (108)$$

$$\lambda = -1: \qquad (\boldsymbol{i}+\xi\boldsymbol{j}),\ (\boldsymbol{i}-\xi\boldsymbol{j}),\ (\boldsymbol{I}+\xi\boldsymbol{J}),\ (\boldsymbol{I}-\xi\boldsymbol{J})\ . \quad (109)$$

Thus it is clearly seen that in case of the particle in homogeneous magnetic field the states corresponding to the longitudinal polarization of octonic oscillator are the eigenstates of $\hat{\pi}_z$ operator with eigenvalue $\lambda = +1$ and for transversal (perpendicular) polarization $\lambda = -1$.

Thus the $\hat{\pi}_z$ operator can be considered as the operator of polarization. At that projection operators

$$\hat{\pi}_+ = \frac{1+\hat{\pi}_z}{2},\quad \hat{\pi}_- = \frac{1-\hat{\pi}_z}{2}, \quad (110)$$

extract from the arbitrary wave function the states with longitudinal and perpendicular polarization of octonic oscillator relatively the Z axis.

**12. Octospinor representation of the operators and wave functions**

The octonic wave function can be written in different representations. In this section we consider general principles for constructing spinor-type representations. In particularly we introduce the eight-dimensional octospinors, which in an explicit form describe a new spatial value characterizing the particle state namely the polarization of octonic oscillator.

In sections 8 and 10 it was shown that the octonic wave function and octonic operators can be written in spinor and bispinor representations. In the octonic quantum mechanics the basis functions of these representations are octons with definite spatial structure. The concrete form of the structure depends on the set of commuting operators, which correspond to diagonal matrices in this representation. The dimension of the representation depends on the number of independent operators in this set, which can not be expressed in terms of each other (hereinafter we will use the term "the fundamental set of operators" for these operators). For example, the two-dimensional spinor representation is based on one operator $\hat{\boldsymbol{K}}$, and the four-dimensional bispinor representation is based on two independent operators $\hat{\boldsymbol{K}}$ and $\hat{\boldsymbol{E}}$.

But the octonic space is eight-dimensional. Therefore to define an unambiguous basis of some representation we have to take eight linearly independent octons. This means that the representation has to be eight-dimensional, and we have to take three operators as the fundamental set. If the number of basis functions is less than eight, these basis functions have underdetermined degrees of freedom. In sections 8 and 10 we have seen the appearance of such degrees of freedom in basis functions of spinor and bispinor representations.

Note, that in the octonic quantum mechanics the octonic wave function contains all the information about particle state in contrast to spinor and bispinor wave functions. But in many physical problems we are interested only in some values, for example, spin projection. For these problems it is more convenient to use the spinor-type representations, in which a quantum system is considered only in respect to the set of states and their changing. In these representations a concrete form of the wave functions and operators is not important, therefore operators can be represented as matrices and wave functions as vector-columns.

Let us consider general principles of correct spinor-type representations constructing. To define a representation it is enough to specify the set of octonic functions which form the representation basis. At that the basis functions are eigenfunctions of every fundamental operator simultaneously.

Except the fundamental operators, the operators which convert any basis function of the representation into another basis function are also very important. These operators have matrix form in considered representation. For example, in the spinor representation octonic operators $\hat{I}$, $\hat{J}$, $\hat{K}$ have matrix form (the Pauli matrices), and operators $\hat{i}$, $\hat{j}$, $\hat{k}$, $\hat{E}$ do not have such a form. Among the operators, which have matrix form, one can select the minimal set of operators (every operator which has matrix form can be expressed in terms of operators from this set). Hereinafter we will consider only minimal sets, which contain the fundamental set of operators (the minimal set can always be selected under this requirement).

Then every spinor-type representation of the octonic wave function can be built according to the following algorithm.

1. Define a set of operators, which would have the matrix form in the representation.
2. Define the fundamental set of operators, which would correspond to diagonal matrices in the representation.
3. Select the minimal set of operators.
4. Build the basis functions of the representation as eigenfunctions of each operator from the fundamental set simultaneously.
5. Select the concrete matrix form for each operator from the minimal set. This would unambiguously define the basis functions of the representation.

Let us apply the considered algorithm to construct eight-dimensional octospinor representation of the octonic wave function. Note that among the octonic basis operators one can select at most two different commuting operators, for example, $\hat{K}$ and $\hat{E}$. But to define the basis functions of the representation unambiguously it is necessary to select 3 operators. This problem can be solved with the help of octonic operators of spatial rotations $\hat{\pi}_x$, $\hat{\pi}_y$, $\hat{\pi}_z$ and reflection operators $\hat{R}_x$, $\hat{R}_y$, $\hat{R}_z$, $\hat{R}$. In this case the operator $\hat{\pi}_z$ can be selected as the third operator which commutes with $\hat{K}$ and $\hat{E}$. Thus let us build the representation in which the operators $\hat{K}$, $\hat{E}$ and $\hat{\pi}_z$ would be fundamental.

Let these operators have the following matrix form in octospinor representation:

$$\hat{K} = \begin{pmatrix} 1 & 0 & 0 & 0 & 0 & 0 & 0 & 0 \\ 0 & -1 & 0 & 0 & 0 & 0 & 0 & 0 \\ 0 & 0 & 1 & 0 & 0 & 0 & 0 & 0 \\ 0 & 0 & 0 & -1 & 0 & 0 & 0 & 0 \\ 0 & 0 & 0 & 0 & 1 & 0 & 0 & 0 \\ 0 & 0 & 0 & 0 & 0 & -1 & 0 & 0 \\ 0 & 0 & 0 & 0 & 0 & 0 & 1 & 0 \\ 0 & 0 & 0 & 0 & 0 & 0 & 0 & -1 \end{pmatrix}, \quad \hat{E} = \begin{pmatrix} 1 & 0 & 0 & 0 & 0 & 0 & 0 & 0 \\ 0 & 1 & 0 & 0 & 0 & 0 & 0 & 0 \\ 0 & 0 & 1 & 0 & 0 & 0 & 0 & 0 \\ 0 & 0 & 0 & 1 & 0 & 0 & 0 & 0 \\ 0 & 0 & 0 & 0 & -1 & 0 & 0 & 0 \\ 0 & 0 & 0 & 0 & 0 & -1 & 0 & 0 \\ 0 & 0 & 0 & 0 & 0 & 0 & -1 & 0 \\ 0 & 0 & 0 & 0 & 0 & 0 & 0 & -1 \end{pmatrix}, \quad \hat{\pi}_z = \begin{pmatrix} 1 & 0 & 0 & 0 & 0 & 0 & 0 & 0 \\ 0 & 1 & 0 & 0 & 0 & 0 & 0 & 0 \\ 0 & 0 & -1 & 0 & 0 & 0 & 0 & 0 \\ 0 & 0 & 0 & -1 & 0 & 0 & 0 & 0 \\ 0 & 0 & 0 & 0 & 1 & 0 & 0 & 0 \\ 0 & 0 & 0 & 0 & 0 & 1 & 0 & 0 \\ 0 & 0 & 0 & 0 & 0 & 0 & -1 & 0 \\ 0 & 0 & 0 & 0 & 0 & 0 & 0 & -1 \end{pmatrix}. \quad (111)$$

Then the basis functions of this representation would have the following form:

$$\chi_1 = a_1(1 + K + E + k),$$
$$\chi_2 = a_2(1 - K + E - k),$$
$$\chi_3 = a_3(I + \xi J + i + \xi j),$$
$$\chi_4 = a_4(I - \xi J + i - \xi j), \quad (112)$$
$$\chi_5 = a_5(1 + K - E - k),$$
$$\chi_6 = a_6(1 - K - E + k),$$
$$\chi_7 = a_7(I + \xi J - i - \xi j),$$
$$\chi_8 = a_8(I - \xi J - i + \xi j).$$

Here $a_s$ ($s = 1..8$) are arbitrary constants. Let the operators $\hat{I}$, $\hat{R}_x$ and $\hat{R}$ are the rest operators of the minimal set. In accordance with the rules of commutation and multiplication we will define the matrix form of these operators as follows

$$\hat{I} = \begin{pmatrix} 0 & 0 & 0 & 1 & 0 & 0 & 0 & 0 \\ 0 & 0 & 1 & 0 & 0 & 0 & 0 & 0 \\ 0 & 1 & 0 & 0 & 0 & 0 & 0 & 0 \\ 1 & 0 & 0 & 0 & 0 & 0 & 0 & 0 \\ 0 & 0 & 0 & 0 & 0 & 0 & 0 & 1 \\ 0 & 0 & 0 & 0 & 0 & 0 & 1 & 0 \\ 0 & 0 & 0 & 0 & 0 & 1 & 0 & 0 \\ 0 & 0 & 0 & 0 & 1 & 0 & 0 & 0 \end{pmatrix}, \quad \hat{R}_x = \begin{pmatrix} 0 & 0 & 0 & 0 & 0 & 1 & 0 & 0 \\ 0 & 0 & 0 & 0 & 1 & 0 & 0 & 0 \\ 0 & 0 & 0 & 0 & 0 & 0 & 0 & 1 \\ 0 & 0 & 0 & 0 & 0 & 0 & 1 & 0 \\ 0 & 1 & 0 & 0 & 0 & 0 & 0 & 0 \\ 1 & 0 & 0 & 0 & 0 & 0 & 0 & 0 \\ 0 & 0 & 0 & 1 & 0 & 0 & 0 & 0 \\ 0 & 0 & 1 & 0 & 0 & 0 & 0 & 0 \end{pmatrix}, \quad \hat{R} = \begin{pmatrix} 0 & 0 & 0 & 0 & 1 & 0 & 0 & 0 \\ 0 & 0 & 0 & 0 & 0 & 1 & 0 & 0 \\ 0 & 0 & 0 & 0 & 0 & 0 & 1 & 0 \\ 0 & 0 & 0 & 0 & 0 & 0 & 0 & 1 \\ 1 & 0 & 0 & 0 & 0 & 0 & 0 & 0 \\ 0 & 1 & 0 & 0 & 0 & 0 & 0 & 0 \\ 0 & 0 & 1 & 0 & 0 & 0 & 0 & 0 \\ 0 & 0 & 0 & 1 & 0 & 0 & 0 & 0 \end{pmatrix}. \quad (113)$$

Then for basis functions (112) of the octospinor representation we get

$$a_1 = a_2 = a_3 = a_4 = a_5 = a_6 = a_7 = a_8 \equiv a/4. \quad (114)$$

The basis functions can be written in the following form:

$$\chi_1 = a\left(\frac{1+E}{2}\right)\left(\frac{1+K}{2}\right),$$
$$\chi_2 = a\left(\frac{1+E}{2}\right)\left(\frac{1-K}{2}\right),$$
$$\chi_3 = a\left(\frac{1+E}{2}\right)\left(\frac{I+\xi J}{2}\right),$$
$$\chi_4 = a\left(\frac{1+E}{2}\right)\left(\frac{I-\xi J}{2}\right), \quad (115)$$
$$\chi_5 = a\left(\frac{1-E}{2}\right)\left(\frac{1+K}{2}\right),$$
$$\chi_6 = a\left(\frac{1-E}{2}\right)\left(\frac{1-K}{2}\right),$$
$$\chi_7 = a\left(\frac{1-E}{2}\right)\left(\frac{I+\xi J}{2}\right),$$
$$\chi_8 = a\left(\frac{1-E}{2}\right)\left(\frac{I-\xi J}{2}\right).$$

The constant $a$ should be defined from normalization condition. Comparing (115) with (89), one can see that the octospinor representation allows to separate the states of quantum system with different polarization of octonic oscillator, which are mixed in spinor and bispinor representations. At that the octonic wave function is characterized with new quantum number – polarization of the octonic oscillator.

Thus we have shown that in the octospinor representation, which is based on 3 commuting operators $\hat{K}$, $\hat{E}$ and $\hat{\pi}_z$, the basis functions can be completely defined. They correspond to the states of particle or antiparticle with determined energy, spin projection and polarization of the octonic oscillator.

## 13. Rotation transform of the octonic wave function

The rotation of the octonic wave function $\breve{\psi}$ on the angle $\theta$ around the unit pseudovector $\vec{n}$ is given by octon

$$\breve{U} = \cos(\theta/2) + \xi \sin(\theta/2)\vec{n}. \tag{116}$$

The transformed function $\breve{\psi}'$ is defined by octonic product

$$\breve{\psi}' = \breve{U}^* \breve{\psi}\, \breve{U}, \tag{117}$$

where octon $\breve{U}^*$ is complex conjugated to $\breve{U}$:

$$\breve{U}^* = \cos(\theta/2) - \xi \sin(\theta/2)\vec{n}, \tag{118}$$

therefore

$$\breve{U}^* \breve{U} = 1. \tag{119}$$

The transformed function $\breve{\psi}'$ can be written on the base of (116)-(118) as

$$\begin{aligned}\breve{\psi}' &= \left[\cos(\theta/2) - \xi \sin(\theta/2)\vec{n}\right](\psi_0 + \vec{\psi} + \tilde{\varphi}_0 + \vec{\varphi})\left[\cos(\theta/2) + \xi \sin(\theta/2)\vec{n}\right] = \\ &= \psi_0 + \vec{\psi}\cos\theta + (1-\cos\theta)(\vec{n},\vec{\psi})\vec{n} - \xi \sin\theta[\vec{n},\vec{\psi}] + \\ &\quad + \tilde{\varphi}_0 + \vec{\varphi}\cos\theta + (1-\cos\theta)(\vec{n},\vec{\varphi})\vec{n} - \xi \sin\theta[\vec{n},\vec{\varphi}].\end{aligned} \tag{120}$$

It is clearly seen that rotation (117) does not transform scalar and pseudoscalar parts of the wave function. On the other hand the vector $\vec{\psi}$ and pseudovector $\vec{\varphi}$ of the wave function are rotated on angle $\theta$ around $\vec{n}$ (see appendix 6).

Transformation (117) contradicts to the rotation of spinors and bispinors [25, 26]. Indeed the rotation of spinor (or bispinor) function $\psi$ is written as

$$\psi' = \hat{U}^{-1}\psi, \tag{121}$$

where $\psi'$ is transformed spinor (or bispinor) function. In spinor representation the turn operator has the form

$$\hat{U} = \cos(\theta/2) + \xi \sin(\theta/2)\vec{n}\hat{\vec{\sigma}}, \tag{122}$$

where $\hat{\vec{\sigma}}$ is the vector of Pauli matrices. In bispinor representation this operator has the form

$$\hat{U} = \cos(\theta/2) + \xi \sin(\theta/2)\vec{n}\hat{\vec{\Sigma}}, \tag{123}$$

where $\hat{\vec{\Sigma}}$ is vector of Dirac $\Sigma$ matrices. It is seen that transformation (121) changes the sign of spinor and bispinor wave functions under turn on $2\pi$. On the other hand the octonic rotation on the angle $2\pi$ (117) does not change the wave function. But as it was shown in sections 6 and 9 all these functions should describe the same physical reality.

To clarify this contradiction we will pass on to the bispinor representation in (117). Then the wave functions $\breve{\psi}$, $\breve{\psi}'$ and rotation octons $\breve{U}$, $\breve{U}^*$ will be matrix operators and (117) can be represented as

$$\hat{\psi}' = \hat{U}^* \hat{\psi}\, \hat{U}. \tag{124}$$

Using (119) the operator we can rewrite the expression (124) in the following form:

$$\hat{\psi}' = \hat{U}^{-1} \hat{\psi}\, \hat{U}. \tag{125}$$

It fact the expression (125) is the rule of turn transformation for octonic operators in bispinor representation. This rule coincides with transformation rules for matrix operators in the Dirac theory [27]:

$$\hat{D}' = \hat{U}^{-1} \hat{D} \hat{U}, \tag{126}$$

where $\hat{D}$ and $\hat{D}'$ are some matrix operator (for example $\hat{\Sigma}_x$) before and after rotation and the matrix $\hat{U}$ coincides with matrix obtained from octon (116). Thus the turn transformation for the octonic wave function coincides with transformation for matrices. But the Dirac wave function is a bispinor which is transformed as column but not a matrix. This is the cause of contradiction. The same contradiction is observed also in the Pauli theory. Actually these theories do not take into account the spatial structure of basis functions of spinor and bispinor. This fact leads to the incorrect description of the spatial rotation.

Consequently for the correct description of the wave function rotation they should use the octonic but not spinor representation. On the base of (125) we can propose a matrix form of the wave function, which is transformed correctly. This function is obtained by means of changing unit vectors in octonic wave function by corresponding matrices of bispinor representation. For the arbitrary wave function we get the following $4 \times 4$ matrix:

$$\hat{\psi} = \begin{pmatrix} \psi_0 + \psi_z + \varphi_0 + \varphi_z & \psi_x - \xi\psi_y + \varphi_x - \xi\varphi_y & 0 & 0 \\ \psi_x + \xi\psi_y + \varphi_x + \xi\varphi_y & \psi_0 - \psi_z + \varphi_0 - \varphi_z & 0 & 0 \\ 0 & 0 & \psi_0 - \psi_z - \varphi_0 + \varphi_z & -\psi_x + \xi\psi_y + \varphi_x + \xi\varphi_y \\ 0 & 0 & -\psi_x - \xi\psi_y + \varphi_x - \xi\varphi_y & \psi_0 + \psi_z - \varphi_0 - \varphi_z \end{pmatrix}. \tag{127}$$

The spatial turn of the wave function (127) is realized on the base of (125).

## 14. Lorentz transformation of the octonic wave function

The Lorentz transformation of the octonic wave function $\breve{\psi}$ is given by

$$\breve{S} = \operatorname{ch}(u/2) - \operatorname{sh}(u/2)\vec{n}, \tag{128}$$

where $\operatorname{th} u = v/c$, $v$ is velocity of motion along the unit vector $\vec{n}$. The transformed function $\breve{\psi}'$ is defined by octonic product

$$\breve{\psi}' = \breve{S}\, \breve{\psi}\, \breve{S}. \tag{129}$$

The transformed function $\breve{\psi}'$ can be written on the base of (128)-(129) as

$$\begin{aligned}\breve{\psi}' &= \left[\operatorname{ch}(u/2) - \operatorname{sh}(u/2)\vec{n}\right](\psi_0 + \vec{\psi} + \tilde{\varphi}_0 + \vec{\varphi})\left[\operatorname{ch}(u/2) - \operatorname{sh}(u/2)\vec{n}\right] = \\ &= \psi_0 \operatorname{ch} u + \vec{\psi} - \psi_0 \vec{n} \operatorname{sh} u - (\vec{n}, \vec{\psi})\operatorname{sh} u - (1 - \operatorname{ch} u)(\vec{n}, \vec{\psi})\vec{n} + \\ &\quad + \tilde{\varphi}_0 \operatorname{ch} u + \vec{\varphi} - \tilde{\varphi}_0 \vec{n}\operatorname{sh} u - (\vec{n}, \vec{\varphi})\operatorname{sh} u - (1 - \operatorname{ch} u)(\vec{n}, \vec{\varphi})\vec{n}.\end{aligned} \tag{130}$$

The components of transformed octonic function have the following structure:

$$\psi'_0 = \psi_0 \operatorname{ch} u - (\vec{n}, \vec{\psi}) \operatorname{sh} u,$$
$$\tilde{\varphi}'_0 = \tilde{\varphi}_0 \operatorname{ch} u - (\vec{n}, \vec{\varphi}) \operatorname{sh} u,$$
$$\vec{\psi}' = \vec{\psi} - \psi_0 \vec{n} \operatorname{sh} u - (1 - \operatorname{ch} u)(\vec{n}, \vec{\psi}) \vec{n}, \quad (131)$$
$$\vec{\varphi}' = \vec{\varphi} - \tilde{\varphi}_0 \vec{n} \operatorname{sh} u - (1 - \operatorname{ch} u)(\vec{n}, \vec{\varphi}) \vec{n}.$$

It is seen that the longitudinal and transversal (relatively direction $\vec{n}$) components of the wave function are transformed as coordinates and time.

## 15. Equations for octonic quantum fields

Formally the octonic relativistic second-order equation (6) corresponding to the Einstein relation between energy and momentum

$$\left(\frac{1}{c}\frac{\partial}{\partial t} - \vec{\nabla}\right)\left(\frac{1}{c}\frac{\partial}{\partial t} + \vec{\nabla}\right)\breve{\psi} = -\frac{m^2 c^2}{\hbar^2}\breve{\psi} \quad (132)$$

is similar to the octonic wave equation for the potentials of electromagnetic field [24]. Therefore in the relativistic octonic quantum mechanics we can define some quantum fields, which will satisfy the first-order equations analogous to the Maxwell equations.

Indeed, by means of operator of spatial inversion $\hat{R}$, which anticommutes with operator $\vec{\nabla}$ the equation (132) can be represented in the form

$$\left(\frac{1}{c}\frac{\partial}{\partial t} - \vec{\nabla} - \xi\frac{mc}{\hbar}\hat{R}\right)\left(\frac{1}{c}\frac{\partial}{\partial t} + \vec{\nabla} + \xi\frac{mc}{\hbar}\hat{R}\right)\breve{\psi} = 0, \quad (133)$$

where the left part consists of the product of two octonic operators. The representation (133) allows to define quantum fields and to obtain the system of first-order equations. This procedure is absolutely analogous to the procedure of obtaining the Maxwell equations in electrodynamics [24]. Let consider the sequential action of operators in (133). After the action of the first operator we obtain the expression

$$\left(\frac{1}{c}\frac{\partial}{\partial t} + \vec{\nabla} + \xi\frac{mc}{\hbar}\hat{R}\right)(\psi_0 + \vec{\psi} + \tilde{\varphi}_0 + \vec{\varphi}) = \frac{1}{c}\frac{\partial \psi_0}{\partial t} + \frac{1}{c}\frac{\partial \vec{\psi}}{\partial t} + \frac{1}{c}\frac{\partial \tilde{\varphi}_0}{\partial t} + \frac{1}{c}\frac{\partial \vec{\varphi}}{\partial t} +$$
$$+ \vec{\nabla}\psi_0 + (\vec{\nabla},\vec{\psi}) + [\vec{\nabla},\vec{\psi}] + \vec{\nabla}\tilde{\varphi}_0 + (\vec{\nabla},\vec{\varphi}) + [\vec{\nabla},\vec{\varphi}] + \xi\frac{mc}{\hbar}\psi_0 - \xi\frac{mc}{\hbar}\vec{\psi} - \xi\frac{mc}{\hbar}\tilde{\varphi}_0 + \xi\frac{mc}{\hbar}\vec{\varphi}, \quad (134)$$

which allows to define quantum fields on the base of the wave function. We will indicate these fields by index $\psi$:

$$e_\psi = \frac{1}{c}\frac{\partial \psi_0}{\partial t} + (\vec{\nabla},\vec{\psi}) + \xi\frac{mc}{\hbar}\psi_0, \quad (135)$$

$$\vec{E}_\psi = -\vec{\nabla}\psi_0 - \frac{1}{c}\frac{\partial \vec{\psi}}{\partial t} + \xi\frac{mc}{\hbar}\vec{\psi} - [\vec{\nabla},\vec{\varphi}], \quad (136)$$

$$\tilde{h}_\psi = \frac{\xi}{c}\frac{\partial \tilde{\varphi}_0}{\partial t} + \xi(\vec{\nabla},\vec{\varphi}) + \frac{mc}{\hbar}\tilde{\varphi}_0, \quad (137)$$

$$\vec{\tilde{H}}_\psi = -\xi[\vec{\nabla},\vec{\psi}] - \xi\vec{\nabla}\tilde{\varphi}_0 - \frac{\xi}{c}\frac{\partial \vec{\varphi}}{\partial t} + \frac{mc}{\hbar}\vec{\varphi}. \quad (138)$$

Here $e_\psi$ is a scalar field, $\vec{E}_\psi$ is a vector field, $\tilde{h}_\psi$ is a pseudoscalar field, $\vec{\tilde{H}}_\psi$ is a pseudovector field. Then from (133) we get the equation

$$\left(\frac{1}{c}\frac{\partial}{\partial t} - \vec{\nabla} - \xi\frac{mc}{\hbar}\hat{R}\right)(e_\psi - \vec{E}_\psi - \xi\tilde{h}_\psi + \xi\vec{\tilde{H}}_\psi) = 0, \quad (139)$$

which leads us to the system of first-order equations analogous to the Maxwell equations:

$$\left(\vec{\nabla}, \vec{E}_\psi\right) = -\frac{1}{c}\frac{\partial e_\psi}{\partial t} + \xi\frac{mc}{\hbar}e_\psi \qquad \text{is scalar part,} \tag{140}$$

$$\left(\vec{\nabla}, \vec{H}_\psi\right) = -\frac{1}{c}\frac{\partial \tilde{h}_\psi}{\partial t} - \xi\frac{mc}{\hbar}\tilde{h}_\psi \qquad \text{is pseudoscalar part,} \tag{141}$$

$$\left[\vec{\nabla}, \vec{H}_\psi\right] = \frac{\xi}{c}\frac{\partial \vec{E}_\psi}{\partial t} + \xi\vec{\nabla}e_\psi - \frac{mc}{\hbar}\vec{E}_\psi \qquad \text{is vector part,} \tag{142}$$

$$\left[\vec{\nabla}, \vec{E}_\psi\right] = -\frac{\xi}{c}\frac{\partial \vec{H}_\psi}{\partial t} - \xi\vec{\nabla}\tilde{h}_\psi - \frac{mc}{\hbar}\vec{H}_\psi \qquad \text{is pseudovector part.} \tag{143}$$

This system is absolutely equivalent to the equation (133).

By analogy with electrodynamics we can construct two-vector first-order equations only for fields $\vec{E}_\psi$ and $\vec{H}_\psi$. The conditions $e_\psi = 0$ and $\tilde{h}_\psi = 0$ lead to the following gauge relations for the wave function:

$$\frac{1}{c}\frac{\partial \psi_0}{\partial t} + \left(\vec{\nabla}, \vec{\psi}\right) + \xi\frac{mc}{\hbar}\psi_0 = 0, \tag{144}$$

$$\frac{1}{c}\frac{\partial \tilde{\varphi}_0}{\partial t} + \left(\vec{\nabla}, \vec{\varphi}\right) - \xi\frac{mc}{\hbar}\tilde{\varphi}_0 = 0. \tag{145}$$

Under the gauge conditions (144)-(145) the equations for quantum fields can be written as

$$\left(\vec{\nabla}, \vec{E}_\psi\right) = 0, \tag{146}$$

$$\left(\vec{\nabla}, \vec{H}_\psi\right) = 0, \tag{147}$$

$$\left[\vec{\nabla}, \vec{H}_\psi\right] = \frac{\xi}{c}\frac{\partial \vec{E}_\psi}{\partial t} - \frac{mc}{\hbar}\vec{E}_\psi, \tag{148}$$

$$\left[\vec{\nabla}, \vec{E}_\psi\right] = -\frac{\xi}{c}\frac{\partial \vec{H}_\psi}{\partial t} - \frac{mc}{\hbar}\vec{H}_\psi. \tag{149}$$

Note that if we take the mass equal to zero and choose the wave function as the four component potential of electromagnetic field then the system (146)-(149) will coincide with the Maxwell equations, and the condition (144) will coincide with the Lorentz gauge.

However for the description of $\vec{E}_\psi$ and $\vec{H}_\psi$ on the base of (146)-(149) the four-component wave function is quite enough. Indeed the solution of equations (146) and (147) can be represented as

$$\vec{E}_\psi = \left[\vec{\nabla}, \vec{\varphi}'\right], \tag{150}$$

$$\vec{H}_\psi = \left[\vec{\nabla}, \vec{\psi}'\right], \tag{151}$$

where $\vec{\psi}'$ and $\vec{\varphi}'$ are some vector and pseudovector functions. Substituting (150) and (151) into equations (148) and (149) we get

$$\left[\vec{\nabla}, \left(\frac{\xi}{c}\frac{\partial \vec{\varphi}'}{\partial t} - \frac{mc}{\hbar}\vec{\varphi}' - \left[\vec{\nabla}, \vec{\psi}'\right]\right)\right] = 0, \tag{152}$$

$$\left[\vec{\nabla}, \left(\frac{\xi}{c}\frac{\partial \vec{\psi}'}{\partial t} + \frac{mc}{\hbar}\vec{\psi}' + \xi\left[\vec{\nabla}, \vec{\varphi}'\right]\right)\right] = 0. \tag{153}$$

The solutions of the equations (152)-(153) have the form

$$\frac{\xi}{c}\frac{\partial \vec{\tilde{\varphi}}'}{\partial t} - \frac{mc}{\hbar}\vec{\tilde{\varphi}}' - \left[\vec{\nabla},\vec{\psi}'\right] = \vec{\nabla}\tilde{\varphi}_0', \tag{154}$$

$$\frac{\xi}{c}\frac{\partial \vec{\psi}'}{\partial t} + \frac{mc}{\hbar}\vec{\psi}' + \xi\left[\vec{\nabla},\vec{\tilde{\varphi}}'\right] = \vec{\nabla}\psi_0', \tag{155}$$

where $\psi_0'$ and $\tilde{\varphi}_0'$ are arbitrary scalar and pseudoscalar functions. So the quantum fields can be expressed through the functions $\psi_0'$ and $\vec{\psi}'$:

$$\vec{E}_\psi = \left[\vec{\nabla},\vec{\tilde{\varphi}}'\right] = -\xi\vec{\nabla}\psi_0' - \frac{1}{c}\frac{\partial \vec{\psi}'}{\partial t} + \xi\frac{mc}{\hbar}\vec{\psi}', \tag{156}$$

$$\vec{H}_\psi = \left[\vec{\nabla},\vec{\psi}'\right]. \tag{157}$$

Analogously the quantum fields can be expressed through the functions $\tilde{\varphi}_0'$ and $\vec{\tilde{\varphi}}'$:

$$\vec{E}_\psi = \left[\vec{\nabla},\vec{\tilde{\varphi}}'\right], \tag{158}$$

$$\vec{H}_\psi = \left[\vec{\nabla},\vec{\psi}'\right] = -\vec{\nabla}\tilde{\varphi}_0' + \frac{\xi}{c}\frac{\partial \vec{\tilde{\varphi}}'}{\partial t} - \frac{mc}{\hbar}\vec{\tilde{\varphi}}'. \tag{159}$$

In both cases for fields' description the four-component wave function is enough. Let consider the fields' definitions (156)-(157). We can find the maximal indetermination of the functions $\psi_0'$ and $\vec{\psi}'$ definition. Since the fields should remain unchanged we get the following conditions:

$$\left[\vec{\nabla},\vec{\psi}''\right] = 0, \tag{160}$$

$$-\xi\vec{\nabla}\psi_0'' - \frac{1}{c}\frac{\partial \vec{\psi}''}{\partial t} + \xi\frac{mc}{\hbar}\vec{\psi}'' = 0. \tag{161}$$

The solution of the equation (160) can be written in form

$$\vec{\psi}'' = \vec{\nabla}F, \tag{162}$$

where $F$ is arbitrary scalar function. Then from (161) we get

$$\psi_0'' = \frac{\xi}{c}\frac{\partial F}{\partial t} + \frac{mc}{\hbar}F. \tag{163}$$

Thus the maximal indetermination of the functions $\psi_0'$ and $\vec{\psi}'$ is defined by one scalar function $F$. Consequently, considered potentials should satisfy one scalar gauge condition

$$\left(\vec{\nabla},\vec{\psi}'\right) + \frac{1}{c}\frac{\partial \psi_0'}{\partial t} + \xi\frac{mc}{\hbar}\psi_0' = 0. \tag{164}$$

So, postulating the equations (146)-(149) for two-vector field leads to the essential restriction of the class of wave functions contrary to the second order equation (132).

The system of equations (140)-(143) can be generalized for the particle in an external electromagnetic field. In this case we have to change operators in (133) by

$$\frac{\partial}{\partial t} \to \frac{\partial}{\partial t} + \frac{\xi e}{\hbar}\Phi, \qquad \vec{\nabla} \to \vec{\nabla} - \frac{\xi e}{\hbar c}\vec{A}. \tag{165}$$

Then we obtain the equation

$$\left(\frac{1}{c}\frac{\partial}{\partial t} + \frac{\xi e}{\hbar c}\Phi - \vec{\nabla} + \frac{\xi e}{\hbar c}\vec{A} - \xi\frac{mc}{\hbar}\hat{R}\right)\left(\frac{1}{c}\frac{\partial}{\partial t} + \frac{\xi e}{\hbar c}\Phi + \vec{\nabla} - \frac{\xi e}{\hbar c}\vec{A} + \xi\frac{mc}{\hbar}\hat{R}\right)\breve{\psi} = 0. \tag{166}$$

After the action of the first operator we have

$$\left(\frac{1}{c}\frac{\partial}{\partial t}+\frac{\xi e}{\hbar c}\Phi+\vec{\nabla}-\frac{\xi e}{\hbar c}\vec{A}+\xi\frac{mc}{\hbar}\hat{R}\right)(\psi_0+\vec{\psi}+\tilde{\varphi}_0+\vec{\varphi})=\frac{1}{c}\frac{\partial\psi_0}{\partial t}+\frac{1}{c}\frac{\partial\vec{\psi}}{\partial t}+\frac{1}{c}\frac{\partial\tilde{\varphi}_0}{\partial t}+\frac{1}{c}\frac{\partial\vec{\varphi}}{\partial t}+$$

$$+\frac{\xi e}{\hbar c}\Phi\psi_0+\frac{\xi e}{\hbar c}\Phi\vec{\psi}+\frac{\xi e}{\hbar c}\Phi\tilde{\varphi}_0+\frac{\xi e}{\hbar c}\Phi\vec{\varphi}+\vec{\nabla}\psi_0+(\vec{\nabla},\vec{\psi})+\left[\vec{\nabla},\vec{\psi}\right]+\vec{\nabla}\tilde{\varphi}_0+(\vec{\nabla},\vec{\varphi})+\left[\vec{\nabla},\vec{\varphi}\right]- \quad (167)$$

$$-\frac{\xi e}{\hbar c}\vec{A}\psi_0-\frac{\xi e}{\hbar c}(\vec{A},\vec{\psi})-\frac{\xi e}{\hbar c}\left[\vec{A},\vec{\psi}\right]-\frac{\xi e}{\hbar c}\vec{A}\tilde{\varphi}_0-\frac{\xi e}{\hbar c}(\vec{A},\vec{\varphi})-\frac{\xi e}{\hbar c}\left[\vec{A},\vec{\varphi}\right]+$$

$$+\xi\frac{mc}{\hbar}\psi_0-\xi\frac{mc}{\hbar}\vec{\psi}-\xi\frac{mc}{\hbar}\tilde{\varphi}_0+\xi\frac{mc}{\hbar}\vec{\varphi}.$$

The quantum fields can be defined as

$$e_\psi=\frac{1}{c}\frac{\partial\psi_0}{\partial t}+(\vec{\nabla},\vec{\psi})+\xi\frac{mc}{\hbar}\psi_0+\frac{\xi e}{\hbar c}\Phi\psi_0-\frac{\xi e}{\hbar c}(\vec{A},\vec{\psi}), \quad (168)$$

$$\vec{E}_\psi=-\vec{\nabla}\psi_0-\frac{1}{c}\frac{\partial\vec{\psi}}{\partial t}+\xi\frac{mc}{\hbar}\vec{\psi}-\left[\vec{\nabla},\vec{\varphi}\right]-\frac{\xi e}{\hbar c}\Phi\vec{\psi}+\frac{\xi e}{\hbar c}\vec{A}\psi_0+\frac{\xi e}{\hbar c}\left[\vec{A},\vec{\varphi}\right], \quad (169)$$

$$\tilde{h}_\psi=\frac{\xi}{c}\frac{\partial\tilde{\varphi}_0}{\partial t}+\xi(\vec{\nabla},\vec{\varphi})+\frac{mc}{\hbar}\tilde{\varphi}_0-\frac{e}{\hbar c}\Phi\tilde{\varphi}_0+\frac{e}{\hbar c}(\vec{A},\vec{\varphi}), \quad (170)$$

$$\vec{H}_\psi=-\xi\left[\vec{\nabla},\vec{\psi}\right]-\xi\vec{\nabla}\tilde{\varphi}_0-\frac{\xi}{c}\frac{\partial\vec{\varphi}}{\partial t}+\frac{mc}{\hbar}\vec{\varphi}+\frac{e}{\hbar c}\Phi\vec{\varphi}-\frac{e}{\hbar c}\left[\vec{A},\vec{\psi}\right]-\frac{e}{\hbar c}\vec{A}\tilde{\varphi}_0. \quad (171)$$

Then equation (166) for the quantum fields takes the form

$$\left(\frac{1}{c}\frac{\partial}{\partial t}+\frac{\xi e}{c\hbar}\Phi-\vec{\nabla}+\frac{\xi e}{\hbar c}\vec{A}-\xi\frac{mc}{\hbar}\hat{R}\right)(e_\psi-\vec{E}_\psi-\xi\tilde{h}_\psi+\xi\vec{H}_\psi)=0. \quad (172)$$

Performing the octonic multiplication we obtain

$$\left(\frac{1}{c}\frac{\partial}{\partial t}+\frac{\xi e}{\hbar c}\Phi-\vec{\nabla}+\frac{\xi e}{\hbar c}\vec{A}-\xi\frac{mc}{\hbar}\hat{R}\right)(e_\psi-\vec{E}_\psi-\xi\tilde{h}_\psi+\xi\vec{H}_\psi)=\frac{1}{c}\frac{\partial e_\psi}{\partial t}-\frac{1}{c}\frac{\partial\vec{E}_\psi}{\partial t}-\frac{\xi}{c}\frac{\partial\tilde{h}_\psi}{\partial t}+\frac{\xi}{c}\frac{\partial\vec{H}_\psi}{\partial t}+$$

$$+\frac{\xi e}{\hbar c}\Phi e_\psi-\frac{\xi e}{\hbar c}\Phi\vec{E}_\psi+\frac{e}{\hbar c}\Phi\tilde{h}_\psi-\frac{e}{\hbar c}\Phi\vec{H}_\psi-\vec{\nabla}e_\psi+(\vec{\nabla},\vec{E}_\psi)+\left[\vec{\nabla},\vec{E}_\psi\right]+\xi\vec{\nabla}\tilde{h}_\psi-\xi(\vec{\nabla},\vec{H}_\psi)-\xi\left[\vec{\nabla},\vec{H}_\psi\right]+ \quad (173)$$

$$+\frac{\xi e}{\hbar c}\vec{A}e_\psi-\frac{\xi e}{\hbar c}(\vec{A},\vec{E}_\psi)-\frac{\xi e}{\hbar c}\left[\vec{A},\vec{E}_\psi\right]+\frac{e}{\hbar c}\vec{A}\tilde{h}_\psi-\frac{e}{\hbar c}(\vec{A},\vec{H}_\psi)-\frac{e}{\hbar c}\left[\vec{A},\vec{H}_\psi\right]-$$

$$-\xi\frac{mc}{\hbar}e_\psi-\xi\frac{mc}{\hbar}\vec{E}_\psi+\frac{mc}{\hbar}\tilde{h}_\psi+\frac{mc}{\hbar}\vec{H}_\psi=0.$$

Then the equations for quantum fields can be written as

$$(\vec{\nabla},\vec{E}_\psi)=-\frac{1}{c}\frac{\partial e_\psi}{\partial t}-\frac{\xi e}{\hbar c}\Phi e_\psi+\frac{\xi e}{\hbar c}(\vec{A},\vec{E}_\psi)+\xi\frac{mc}{\hbar}e_\psi, \quad (174)$$

$$\left[\vec{\nabla},\vec{H}_\psi\right]=\frac{\xi}{c}\frac{\partial\vec{E}_\psi}{\partial t}-\frac{e}{\hbar c}\Phi\vec{E}_\psi+\xi\vec{\nabla}e_\psi+\frac{e}{\hbar c}\vec{A}e_\psi+\frac{\xi e}{\hbar c}\left[\vec{A},\vec{H}_\psi\right]-\frac{mc}{\hbar}\vec{E}_\psi, \quad (175)$$

$$(\vec{\nabla},\vec{H}_\psi)=-\frac{1}{c}\frac{\partial\tilde{h}_\psi}{\partial t}-\frac{\xi e}{\hbar c}\Phi\tilde{h}_\psi+\frac{\xi e}{\hbar c}(\vec{A},\vec{H}_\psi)-\xi\frac{mc}{\hbar}\tilde{h}_\psi, \quad (176)$$

$$\left[\vec{\nabla},\vec{E}_\psi\right]=-\frac{\xi}{c}\frac{\partial\vec{H}_\psi}{\partial t}+\frac{e}{\hbar c}\Phi\vec{H}_\psi-\xi\vec{\nabla}\tilde{h}_\psi+\frac{\xi e}{\hbar c}\left[\vec{A},\vec{E}_\psi\right]-\frac{e}{\hbar c}\vec{A}\tilde{h}_\psi-\frac{mc}{\hbar}\vec{H}_\psi. \quad (177)$$

The system (174)-(177) is absolutely equivalent to the octonic equation (166). Analogously the system (146)-(149) for two-vector fields also can be generalized for the case of particle in an external electromagnetic field.

Note that on the base of systems (140)-(143), (146)-(149), (174)-(177) one can obtain the quadratic forms analogous to the relations between energy and momentum as well as to Lorenz invariants of electromagnetic field in electrodynamics [24].

**16. Octonic first-order equations**

As it was shown in sections 7 and 9 the spin interaction of the particle with electromagnetic field is described in the frames of octonic second-order equation. In contrast to the Dirac theory the terms describing the interaction of spin with electric and magnetic fields are appeared in the octonic second-order equation as a result of octonic multiplication without attraction of the first-order equation. However in octonic quantum mechanics we also can construct the Diracs'-like first-order equations.

Let turn to octonic equation (132):

$$\left(\frac{1}{c}\frac{\partial}{\partial t} - \vec{\nabla}\right)\left(\frac{1}{c}\frac{\partial}{\partial t} + \vec{\nabla}\right)\breve{\psi} = -\frac{m^2 c^2}{\hbar^2}\breve{\psi}. \tag{178}$$

In this equation we can formally denote the result of action of one of operators on function $\breve{\psi}$ as some new octonic function $\breve{W}$:

$$\left(\frac{1}{c}\frac{\partial}{\partial t} + \vec{\nabla}\right)\breve{\psi} = -\frac{mc}{\hbar}\breve{W}. \tag{179}$$

Then the second-order equation (178) is equivalent to the system of two first-order equations:

$$\begin{cases} \left(\frac{1}{c}\frac{\partial}{\partial t} + \vec{\nabla}\right)\breve{\psi} = -\frac{mc}{\hbar}\breve{W}, \\ \left(\frac{1}{c}\frac{\partial}{\partial t} - \vec{\nabla}\right)\breve{W} = \frac{mc}{\hbar}\breve{\psi}. \end{cases} \tag{180}$$

Acting on the second equation of (180) by the operator of spatial inversion $\hat{R}$ we get

$$\begin{cases} \left(\frac{1}{c}\frac{\partial}{\partial t} + \vec{\nabla}\right)\breve{\psi} = \frac{mc}{\hbar}\left(-\breve{W}\right), \\ \left(\frac{1}{c}\frac{\partial}{\partial t} + \vec{\nabla}\right)\hat{R}\breve{W} = \frac{mc}{\hbar}\hat{R}\breve{\psi}. \end{cases} \tag{181}$$

The equations of the system (181) on some conditions can be absolutely equivalent. For that functions $\breve{W}$ and $\breve{\psi}$ should satisfy the following relations:

$$\begin{cases} \breve{\psi} = \eta\hat{R}\breve{W}, \\ -\breve{W} = \eta\hat{R}\breve{\psi}. \end{cases} \tag{182}$$

where $\eta$ is some constant. Acting on the second equality of (182) by operator $\hat{R}$ and performing simple transformation we obtain that $\eta^2 = -1$ and consequently $\eta = \pm\xi$. So if the accessory function $\breve{W}$ satisfies the condition

$$\breve{W} = \pm\xi\hat{R}\breve{\psi}, \tag{183}$$

the wave function $\breve{\psi}$ satisfies the first-order equation. The sign in (183) can be chosen arbitrarily. If $\breve{W} = +\xi \hat{R} \breve{\psi}$ the first-order equation is

$$\left( \frac{1}{c} \frac{\partial}{\partial t} + \vec{\nabla} + \xi \frac{mc}{\hbar} \hat{R} \right) \breve{\psi} = 0. \tag{184}$$

Note that we can also act by $\hat{R}$ on the first equation of system (180). Then we get the equation with other sign before the gradient operator. Thus the first-order equation can be written in four different forms:

$$\left( \frac{1}{c} \frac{\partial}{\partial t} + \vec{\nabla} + \xi \frac{mc}{\hbar} \hat{R} \right) \breve{\psi} = 0, \tag{185}$$

$$\left( \frac{1}{c} \frac{\partial}{\partial t} + \vec{\nabla} - \xi \frac{mc}{\hbar} \hat{R} \right) \breve{\psi} = 0, \tag{186}$$

$$\left( \frac{1}{c} \frac{\partial}{\partial t} - \vec{\nabla} + \xi \frac{mc}{\hbar} \hat{R} \right) \breve{\psi} = 0, \tag{187}$$

$$\left( \frac{1}{c} \frac{\partial}{\partial t} - \vec{\nabla} - \xi \frac{mc}{\hbar} \hat{R} \right) \breve{\psi} = 0. \tag{188}$$

If we search solutions of these equations as the plane waves then each of these equations leads to right relativistic relations between energy and momentum. Indeed, for example, the equation (185) can be represented in component wise form

$$\left( \frac{1}{c} \frac{\partial}{\partial t} + \vec{\nabla} + \xi \frac{mc}{\hbar} \hat{R} \right) \left( \psi_0 + \psi_x \mathbf{i} + \psi_y \mathbf{j} + \psi_z \mathbf{k} + \varphi_0 \mathbf{E} + \varphi_x \mathbf{I} + \varphi_y \mathbf{J} + \varphi_z \mathbf{K} \right) = 0. \tag{189}$$

This equation is equivalent to the system of eight scalar equations

$$\begin{cases} \dfrac{1}{c} \dfrac{\partial \psi_0}{\partial t} + \dfrac{\partial \psi_x}{\partial x} + \dfrac{\partial \psi_y}{\partial y} + \dfrac{\partial \psi_z}{\partial z} + \xi \dfrac{mc}{\hbar} \psi_0 = 0, \\[4pt] \dfrac{1}{c} \dfrac{\partial \psi_x}{\partial t} + \dfrac{\partial \psi_0}{\partial x} + \xi \dfrac{\partial \varphi_z}{\partial y} - \xi \dfrac{\partial \varphi_y}{\partial z} - \xi \dfrac{mc}{\hbar} \psi_x = 0, \\[4pt] \dfrac{1}{c} \dfrac{\partial \psi_y}{\partial t} + \dfrac{\partial \psi_0}{\partial y} + \xi \dfrac{\partial \varphi_x}{\partial z} - \xi \dfrac{\partial \varphi_z}{\partial x} - \xi \dfrac{mc}{\hbar} \psi_y = 0, \\[4pt] \dfrac{1}{c} \dfrac{\partial \psi_z}{\partial t} + \dfrac{\partial \psi_0}{\partial z} + \xi \dfrac{\partial \varphi_z}{\partial y} - \xi \dfrac{\partial \varphi_y}{\partial z} - \xi \dfrac{mc}{\hbar} \psi_z = 0, \\[4pt] \dfrac{1}{c} \dfrac{\partial \varphi_0}{\partial t} + \dfrac{\partial \varphi_x}{\partial x} + \dfrac{\partial \varphi_y}{\partial y} + \dfrac{\partial \varphi_z}{\partial z} + \xi \dfrac{mc}{\hbar} \varphi_0 = 0, \\[4pt] \dfrac{1}{c} \dfrac{\partial \varphi_x}{\partial t} + \dfrac{\partial \varphi_0}{\partial x} + \xi \dfrac{\partial \psi_z}{\partial y} - \xi \dfrac{\partial \psi_y}{\partial z} - \xi \dfrac{mc}{\hbar} \varphi_x = 0, \\[4pt] \dfrac{1}{c} \dfrac{\partial \varphi_y}{\partial t} + \dfrac{\partial \varphi_0}{\partial y} + \xi \dfrac{\partial \psi_x}{\partial z} - \xi \dfrac{\partial \psi_z}{\partial x} - \xi \dfrac{mc}{\hbar} \varphi_y = 0, \\[4pt] \dfrac{1}{c} \dfrac{\partial \varphi_z}{\partial t} + \dfrac{\partial \varphi_0}{\partial z} + \xi \dfrac{\partial \psi_z}{\partial y} - \xi \dfrac{\partial \psi_y}{\partial z} - \xi \dfrac{mc}{\hbar} \varphi_z = 0. \end{cases} \tag{190}$$

If we search the solution as the plane wave

$$\breve{\psi} \sim \exp\left\{ \frac{\xi}{\hbar} \left( -Et + p_x x + p_y y + p_z z \right) \right\}, \tag{191}$$

then the dispersion relation for the system (190) is

$$\left(E^2 - p^2 c^2 - m^2 c^4\right)^4 = 0, \qquad (192)$$

where $p^2 = p_x^2 + p_y^2 + p_z^2$. The roots of equation (192) $E = \pm\sqrt{p^2 c^2 + m^2 c^4}$ are fourthly degenerate.

Note that for obtaining the first-order equation we used only the property of anticommutation between $\hat{R}$ and $\hat{i}, \hat{j}, \hat{k}$ in the operator $\vec{\nabla}$. But the operator $\hat{E}\hat{R}$ has the same commutative properties. So we can propose other first-order equations:

$$\left(\frac{1}{c}\frac{\partial}{\partial t} + \vec{\nabla} + \frac{mc}{\hbar}\hat{E}\hat{R}\right)\breve{\psi} = 0, \qquad (193)$$

$$\left(\frac{1}{c}\frac{\partial}{\partial t} + \vec{\nabla} - \frac{mc}{\hbar}\hat{E}\hat{R}\right)\breve{\psi} = 0, \qquad (194)$$

$$\left(\frac{1}{c}\frac{\partial}{\partial t} - \vec{\nabla} + \frac{mc}{\hbar}\hat{E}\hat{R}\right)\breve{\psi} = 0, \qquad (195)$$

$$\left(\frac{1}{c}\frac{\partial}{\partial t} - \vec{\nabla} - \frac{mc}{\hbar}\hat{E}\hat{R}\right)\breve{\psi} = 0. \qquad (196)$$

Besides the first-order equations can be written with operators $\xi\hat{R}$ or $\hat{E}\hat{R}$ placed before the time derivative.

Thus the procedure of writing the first-order equations on the base of second-order equation is essentially ambiguous. The equations (185)-(188), (193)-(196) and others cannot be brought to each other by any octonic transformations.

Let us compare obtained first-order equations with the Dirac equation written in symmetric form [25]

$$\left(\hat{\gamma}_0 \frac{1}{c}\frac{\partial}{\partial t} + \hat{\vec{\gamma}} \cdot \vec{\nabla} + \xi \frac{mc}{\hbar}\right)\psi = 0. \qquad (197)$$

Here $\psi$ is the wave function in the form of four-component bispinor, $\hat{\gamma}_s$ ($s = 0, 1, 2, 3$) are the Dirac matrix operators. Taking into account relations between octonic spatial operators and the Dirac matrices as well as octonic structure of basis bispinor functions (see section 10) it is easy to show that the equation (197) realizes the bispinor representation of the octonic equation (185). Indeed multiplying (197) on $\hat{\gamma}_0$ and using octonic representation of matrix operators and octonic representation of basis wave functions of bispinor we can obtain

$$\left(\frac{1}{c}\frac{\partial}{\partial t} + \vec{\nabla} + \xi \frac{mc}{\hbar}\hat{R}\right)\breve{\psi} = 0, \qquad (198)$$

which coincides with the equation (185). So the Dirac equation (197) is equivalent to the octonic first-order equation (185).

There is also the inverse procedure of obtaining the second-order equation analogous to procedure, which is used in the Dirac theory. For example, acting on the equation (185) by operator

$$\left(\frac{1}{c}\frac{\partial}{\partial t} - \vec{\nabla} - \xi \frac{mc}{\hbar}\hat{R}\right), \qquad (199)$$

we get the following equation:

$$\left(\frac{1}{c}\frac{\partial}{\partial t} - \vec{\nabla} - \xi \frac{mc}{\hbar}\hat{R}\right)\left(\frac{1}{c}\frac{\partial}{\partial t} + \vec{\nabla} + \xi \frac{mc}{\hbar}\hat{R}\right)\breve{\psi} = 0. \qquad (200)$$

Using the property of associativity and multiplying operators in the left part of (200) we obtain the second-order equation coinciding formally with the equation (132) or (178). The similar procedure can be specified for any equations (185)-(188), (193)-(196). However we specially emphasize though equation (200) coincides in form with the second-order equation (133), but the class of solutions defined by (200) is considerably narrow since the wave function should simultaneously satisfy the first-order equation.

The first-order equations have the interesting interpretation. Since the operator in the equation (185) coincides with operator, which is used for the quantum fields' definition (134), then in fact the first order equation (185) describes the particles, which do not create the quantum fields $e_\psi$, $\vec{E}_\psi$, $\tilde{h}_\psi$, $\vec{H}_\psi$.

## 17. Discussion

### *The second-order equations*

One of the starting points for constructing the relativistic quantum theory is the Einstein relation between energy and momentum of the particle, which is written as

$$E^2 = \vec{p}^2 c^2 + m^2 c^4. \quad (201)$$

In the relativistic quantum mechanics this relation is written as the operator equation for the wave function

$$\left(\hat{E}^2 - \hat{\vec{p}}^2 c^2\right)\psi = m^2 c^4 \psi. \quad (202)$$

This equation contains scalar operator $\hat{\vec{p}}^2 = \hat{p}_x^2 + \hat{p}_y^2 + \hat{p}_z^2$. The form of operator $\hat{\vec{p}}$ in general is ambiguous. The equation (202) can be realized by means of different operators and wave functions. For example choosing scalar wave function and operator $\hat{\vec{p}}$ represented in the basis of Gibbs unit vectors $\vec{i}$, $\vec{j}$, $\vec{k}$

$$\hat{\vec{p}} = \hat{p}_x \vec{i} + \hat{p}_y \vec{j} + \hat{p}_z \vec{k}, \quad (203)$$

the equation (202) can be led to the well known Klein-Gordon equation describing the spinless particles. Moreover one can consider the following matrix expression for the $\hat{\vec{p}}$ operator

$$\hat{\vec{p}} = \hat{p}_x \hat{\sigma}_x + \hat{p}_y \hat{\sigma}_y + \hat{p}_z \hat{\sigma}_z. \quad (204)$$

where $\hat{\sigma}_x$, $\hat{\sigma}_y$, $\hat{\sigma}_z$ are some matrices. For example the Pauli matrices or the Dirac matrices can be used as $\hat{\sigma}_s$ ($s = \{x, y, z\}$). It is only important that these matrices anticommute and $\sigma_s^2 = 1$, that leads to $\hat{\vec{p}}^2 = \hat{p}_x^2 + \hat{p}_y^2 + \hat{p}_z^2$. There are also others approaches to the operator $\hat{\vec{p}}$ representation by quaternions [14, 18].

From this point of view it is not surprising that proposed octonic wave function and octonic momentum operator in the form

$$\hat{\vec{p}} = \hat{p}_x \boldsymbol{i} + \hat{p}_y \boldsymbol{j} + \hat{p}_z \boldsymbol{k} \quad (205)$$

also realize the right representation of the equation (202). However emphasize that octonic representation of operator $\hat{\vec{p}}$ (205) is distinguished since it allows to conserve the vector interpretation of $\hat{\vec{p}}$ and to ensure the correct space transformation properties.

In octonic algebra the classical Einstein relation for the particle in electromagnetic field can be represented in the form

$$\left(E - e\Phi + \vec{p}c - e\vec{A}\right)\left(E - e\Phi - \vec{p}c + e\vec{A}\right) = m^2 c^4. \tag{206}$$

Really performing the octonic multiplication in the left part of (206) and gathering terms we get

$$E^2 - 2e\Phi E + e^2\Phi^2 - \vec{p}^2 c^2 + 2ec(\vec{p}, \vec{A}) - e^2 \vec{A}^2 = m^2 c^4 \tag{207}$$

or

$$\left(E - e\Phi\right)^2 - \left(\vec{p} - \frac{e}{c}\vec{A}\right)^2 c^2 = m^2 c^4. \tag{208}$$

Thus octonic relations (206) and (208) in classical relativistic mechanics are absolutely equivalent. Note that these relations are valid as in octon's algebra as in Gibbs vectors algebra too.

However in quantum mechanics the situation is cardinally changed. The octonic operator equation obtained from (206) can be represented as

$$\left(\hat{E} - e\Phi + \hat{\vec{p}}c - e\vec{A}\right)\left(\hat{E} - e\Phi - \hat{\vec{p}}c + e\vec{A}\right)\tilde{\psi} = m^2 c^4 \tilde{\psi}, \tag{209}$$

but the equation obtained from (208) can be written as

$$\left(\left(\hat{E} - e\Phi\right)^2 - \left(\hat{\vec{p}} - \frac{e}{c}\vec{A}\right)^2 c^2\right)\tilde{\psi} = m^2 c^4 \tilde{\psi}. \tag{210}$$

The equations (209) and (210) are essentially nonequivalent. Really the equation (209) after octonic multiplying leads us to (30) but equation (210) leads us to the following one:

$$\left[-\Delta + \frac{1}{c^2}\frac{\partial^2}{\partial t^2} + \frac{2\xi e}{\hbar c}\left((\vec{A}, \vec{\nabla}) + \frac{\Phi}{c}\frac{\partial}{\partial t}\right) + \frac{m^2 c^2}{\hbar^2} + \frac{e^2}{\hbar^2 c^2}(A^2 - \Phi^2)\right]\tilde{\psi} - \frac{e}{\hbar c}\vec{H}\tilde{\psi} = 0. \tag{211}$$

This expression is differed from (30) by absence of the term describing the interaction of the particle with electric field. Therefore the representation of the Einstein relaition in the form (206) is more adequate than (208) from the view of description of the particle in electromagnetic field.

Consequently in octonic quantum mechanics the second-order equation correctly describing the interaction of spin with electromagnetic field is obtained directly from the Einstein relation in contrast to the Dirac theory where the first order equation is used for obtaining of this equation.

Also using the algebra of octons allows to get nonrelativistic equation, which correctly describes interaction of spin with magnetic field, directly from the Schrödinger equation. Indeed the Schrödinger Hamiltonian contains the operator $\hat{\vec{p}}^2$, which is modified in the presence of magnetic field as

$$\hat{\vec{p}}^2 \rightarrow \left(\hat{\vec{p}} - \frac{e}{c}\vec{A}\right)^2. \tag{212}$$

The octonic multiplication of operators in (212) leads to appearance of the term, which corresponds to the interaction of spin with magnetic field (see (34)). In the Pauli theoty the equation analogous to (36) is postulated. Accordingly the application of the octon's algebra leads to the adequate description of relativistic and nonrelativistic quantum particles in an external electromagnetic field.

Here we note once again that the equation (202) can be realized by means of two cardinally different algebras. The scalar wave function in combination with operator $\hat{\vec{p}}$ in the form of Gibbs vector describes a particle with spin 0 and the octonic wave function with octonic operator $\hat{\vec{p}}$ describes a particle with spin of 1/2.

*Spin and spatial structure of the octonic wave function*

In octonic quantum mechanics a concept of spin is closely connected with spatial properties of octonic operators and octonic wave function. Indeed first of all the operator of spin projection on the Z axis is the pseudovectorial operator $\hat{K}$ corresponding to the unit pseudovector along this direction. In the second place the interaction of spin with magnetic field is described in the second-order equation (30) by the term $\vec{H}\breve{\psi}$, which is the product of two spatial objects pseudovector $\vec{H}$ and octon $\breve{\psi}$. Moreover as it was shown the fact that quantum system can be in states corresponding to particle and antiparticle is also reflected in spatial structure of the wave function.

However representation of the wave function in the octonic form is not always optimal in the concrete physical tasks. Often the spinor-type representations, which emphasize one or several properties of the quantum system, are more convenient. The transition to the spinor representation means in fact the changing of spatial basis of the octonic wave function, and spinor components constitute the expansion coefficients of the octonic wave function on the new basis.

To conserve all information contained in the octonic wave function when transiting to spinor representation the number of basis vectors should be equal to eight, i.e the number of definite parameters of quantum system in that sort of representation should be equal to three. In particularly, we showed that the octonic wave function can be represented in the form of eight-component octospinor, which reflect three properties of the quantum system namely spin projection, projection on states particle/antiparticle and polarization of octonic oscillator. This octospinor representation contains the same information as containing in the octonic wave function. The bispinor describes only two properties and spinor describes only one.

But the octonic representation of the wave function has one incontestable advantage. The octonic wave function is the spatial object and it leads to clear defined geometrical interpretation. For each state of the quantum system there is the quite defined spatial structure of the wave function. For instance, in the state corresponding to the particle vector and pseudovector components of the wave function are modulo equal codirectional vectors but in the state corresponding to the antiparticle vector and pseudovector components of the wave function are opposite directed. In the state with definite spin projection the wave function has the complicated space-time structure in which vector and pseudovector components perform either spatial oscillating along Z direction in case of longitudinal polarization or rotating in the perpendicular plane in case of transversal polarization of octonic oscillator. We especially emphasize that in the state with transversal polarization vector and pseudovector components of the wave function have the real spatial rotation.

*Quantum fields and first-order equations*

The octonic second-order equation corresponding to the Einstein relation between energy and momentum can be represented in the following form:

$$\left(\hat{E}^2 - \hat{\vec{p}}^2 c^2 - m^2 c^4\right)\breve{\psi} = 0. \tag{213}$$

By analogy with octonic electrodynamics [24] the operator in the left part can be represented as a product of two octonic operators. However it can be made by different ways using various sets of spatial operators. For example, using operators $\hat{\vec{i}}$, $\hat{\vec{j}}$, $\hat{\vec{k}}$ and operator $\hat{R}$ or $\hat{E}\hat{R}$ we can represent the equation (213) in four essentially various forms:

$$\left(\frac{1}{c}\frac{\partial}{\partial t} - \vec{\nabla} - \xi\frac{mc}{\hbar}\hat{R}\right)\left(\frac{1}{c}\frac{\partial}{\partial t} + \vec{\nabla} + \xi\frac{mc}{\hbar}\hat{R}\right)\breve{\psi} = 0, \tag{214}$$

$$\left(\frac{1}{c}\frac{\partial}{\partial t} - \vec{\nabla} - \frac{mc}{\hbar}\hat{E}\hat{R}\right)\left(\frac{1}{c}\frac{\partial}{\partial t} + \vec{\nabla} + \frac{mc}{\hbar}\hat{E}\hat{R}\right)\breve{\psi} = 0, \tag{215}$$

$$\left(\xi\hat{R}\frac{1}{c}\frac{\partial}{\partial t} + \vec{\nabla} + \frac{mc}{\hbar}\right)\left(\xi\hat{R}\frac{1}{c}\frac{\partial}{\partial t} + \vec{\nabla} - \frac{mc}{\hbar}\right)\breve{\psi} = 0, \tag{216}$$

$$\left(\hat{E}\hat{R}\frac{1}{c}\frac{\partial}{\partial t} + \vec{\nabla} - \frac{mc}{\hbar}\right)\left(\hat{E}\hat{R}\frac{1}{c}\frac{\partial}{\partial t} + \vec{\nabla} + \frac{mc}{\hbar}\right)\breve{\psi} = 0. \tag{217}$$

The main requirements for representation of relation (213) in the form of equations (214)-(217) are anticommutating of $\hat{R}$ with operators $\hat{i}, \hat{j}, \hat{k}$ and the condition $\hat{R}^2 = 1$. The signs in the interior of equations (214)-(217) can be arranged in two different ways (see (185)-(188)).

On the base of each equation from (214)-(217) we can define the quantum fields, which describe the quantum system as well as the wave function do (the procedure for fields definition is described in section 15). These fields satisfy different first-order equations and have different spatial symmetry.

Note that on the base of (214)-(217) the expressions for quadratic forms of octonic quantum fields analogous to the relations for energy-momentum and Lorenz invariants of electromagnetic field [24] can be obtained.

Thus we showed that there are several kinds of Maxwell's-like first-order equations, which can be put in accordance to the second order equation (213).

On the other hand as it was shown above (section 16) for some bounded classes of the wave functions the second-order equation can be reduced to one of several first-order equations. Since the equation (213) can be realized by several sets of operators (see (214)-(217)) then at least four types of first order equations can be represented:

$$\left(\frac{1}{c}\frac{\partial}{\partial t} + \vec{\nabla} + \xi\frac{mc}{\hbar}\hat{R}\right)\breve{\psi} = 0, \tag{218}$$

$$\left(\frac{1}{c}\frac{\partial}{\partial t} + \vec{\nabla} + \frac{mc}{\hbar}\hat{E}\hat{R}\right)\breve{\psi} = 0, \tag{219}$$

$$\left(\xi\hat{R}\frac{1}{c}\frac{\partial}{\partial t} + \vec{\nabla} - \frac{mc}{\hbar}\right)\breve{\psi} = 0, \tag{220}$$

$$\left(\hat{E}\hat{R}\frac{1}{c}\frac{\partial}{\partial t} + \vec{\nabla} + \frac{mc}{\hbar}\right)\breve{\psi} = 0. \tag{221}$$

There are several forms of equations (218)-(221), which differ by the sign combination in the interior of operators (see (185)-(188)). In fact the equations (218)-(221) realize the conditions of absence of quantum fields defined by equations (214)-(217).

Note that derivation of the Dirac equation on the base of the correspondence principle [27] also contains the similar ambiguity. Indeed, the Dirac equation [28] is written as

$$\left(\frac{1}{c}\frac{\partial}{\partial t} + \alpha_1\frac{\partial}{\partial x} + \alpha_2\frac{\partial}{\partial y} + \alpha_3\frac{\partial}{\partial z} + \xi\beta\frac{mc}{\hbar}\right)\psi = 0. \tag{222}$$

The conditions imposed on the matrix $\beta$ by the Einstein relation (correspondence principle) have the following form:

$$\alpha_s \beta + \beta \alpha_s = 0, \qquad \beta^2 = 1. \tag{223}$$

However the same operator $\beta$ can be put also in other position. In this case the equation satisfying the correspondence principle can be written as

$$\left( \xi\beta \frac{1}{c} \frac{\partial}{\partial t} + \alpha_1 \frac{\partial}{\partial x} + \alpha_2 \frac{\partial}{\partial y} + \alpha_3 \frac{\partial}{\partial z} + \frac{mc}{\hbar} \right) \psi = 0. \tag{224}$$

Indeed acting with the operator

$$\left( \xi\beta \frac{1}{c} \frac{\partial}{\partial t} + \alpha_1 \frac{\partial}{\partial x} + \alpha_2 \frac{\partial}{\partial y} + \alpha_3 \frac{\partial}{\partial z} - \frac{mc}{\hbar} \right) \tag{225}$$

on the equation (224), we get the right second-order equation. The equation (224) is essentially differed from the equation (222) and has different symmetric properties. Moreover we can use the operator $\gamma_5 \beta$ instead of $\beta$ since it has the same commutative properties. In this case the first-order equation, which satisfies the correspondence principle, can be represented in two different forms:

$$\left( \frac{1}{c} \frac{\partial}{\partial t} + \alpha_1 \frac{\partial}{\partial x} + \alpha_2 \frac{\partial}{\partial y} + \alpha_3 \frac{\partial}{\partial z} + \gamma_5 \beta \frac{mc}{\hbar} \right) \psi = 0, \tag{226}$$

$$\left( \gamma_5 \beta \frac{1}{c} \frac{\partial}{\partial t} + \alpha_1 \frac{\partial}{\partial x} + \alpha_2 \frac{\partial}{\partial y} + \alpha_3 \frac{\partial}{\partial z} + \frac{mc}{\hbar} \right) \psi = 0. \tag{227}$$

The same situation is observed for the Dirac equation written in symmetric form [25]:

$$\left( \gamma_0 \frac{1}{c} \frac{\partial}{\partial t} + \gamma_1 \frac{\partial}{\partial x} + \gamma_2 \frac{\partial}{\partial y} + \gamma_3 \frac{\partial}{\partial z} + \xi \frac{mc}{\hbar} \right) \psi = 0, \tag{228}$$

where all matrix operators $\gamma_s$ anticommute to each other. We can indicate another form, which is also satisfied the correspondence principle:

$$\left( \xi \frac{1}{c} \frac{\partial}{\partial t} + \gamma_1 \frac{\partial}{\partial x} + \gamma_2 \frac{\partial}{\partial y} + \gamma_3 \frac{\partial}{\partial z} + \gamma_0 \frac{mc}{\hbar} \right) \psi = 0. \tag{229}$$

The operator, which leads (88) to the right second-order equation, is

$$\left( \xi \frac{1}{c} \frac{\partial}{\partial t} - \gamma_1 \frac{\partial}{\partial x} - \gamma_2 \frac{\partial}{\partial y} - \gamma_3 \frac{\partial}{\partial z} - \gamma_0 \frac{mc}{\hbar} \right). \tag{230}$$

Thus the reduction of the second-order equation to the system of Maxwell's-like first-order equations or to the single Dirac's-like equation is an ambiguous procedure.

Note that the similar ambiguity of constructing the first-order equation exists also on describing the massless particles (neutrino).

However in octonic quantum mechanics there is no need for the first-order equation. As it was shown the terms describing the interaction between spin and electromagnetic field are appeared in the second-order equation automatically without attracting the first-order equation (contrary to Dirac theory where these terms are appeared only at the transition from first-order equations to the second-order equation).

## 16. Conclusion

Thus, in this paper we proposed the scheme for constructing the relativistic quantum mechanics on the base of octonic spatial operators and octonic wave function. It was shown that the octonic second-order equation corresponding to the Einstein relation between energy and momentum leads to the correct description of the interaction between spin and electromagnetic field.

The relations between the octonic wave function and Pauli's spinors and Dirac's bispinors were established. It was shown that the wave function for the states with definite energy and with definite spin projection on the Z axis have a form of octonic oscillator, which is represented as the combination of oscillations of scalar and pseudoscalar components of the wave function, longitudinal oscillations of Z components and rotation of transversal components in the plane perpendicular to the Z axis.

The eight-component octospinors separating the states of the quantum system with different spin projection, different particle-antiparticle state and different polarization of octonic oscillator were constructed.

We showed that in the frames of octonic quantum mechanics the second-order wave equation can be reduced to the system of the first-order Maxwell's-like equations for the quantum fields. Since the Einstein relation can be realized by various sets of spatial operators that it is possible to introduce the quantum fields of several types with different spatial symmetry.

It was shown that for the special class of wave functions the second-order equation can be reduced to the single octonic first-order equation analogous to the Dirac equation. At that since the Einstein relation can be realized by several sets of spatial operators the several types of such first-order equations having different spatial symmetry were proposed. At the same time it was shown that the Dirac's-like first-order equations describe particles, which do not create quantum fields.

## Acknowledgement

The authors are very thankful to G.V. Mironova for kind assistance and supporting.

## 17. Appendixes

### Appendix 1. Matrix form for operators of octonic basis

The operators of octonic basis can be represented in the following matrix form (see (10)):

$$\hat{i} = \begin{pmatrix} 0 & 1 & 0 & 0 & 0 & 0 & 0 & 0 \\ 1 & 0 & 0 & 0 & 0 & 0 & 0 & 0 \\ 0 & 0 & 0 & 0 & 0 & 0 & 0 & -\xi \\ 0 & 0 & 0 & 0 & 0 & 0 & \xi & 0 \\ 0 & 0 & 0 & 0 & 0 & 1 & 0 & 0 \\ 0 & 0 & 0 & 0 & 1 & 0 & 0 & 0 \\ 0 & 0 & 0 & -\xi & 0 & 0 & 0 & 0 \\ 0 & 0 & \xi & 0 & 0 & 0 & 0 & 0 \end{pmatrix}, \quad \hat{j} = \begin{pmatrix} 0 & 0 & 1 & 0 & 0 & 0 & 0 & 0 \\ 0 & 0 & 0 & 0 & 0 & 0 & 0 & \xi \\ 1 & 0 & 0 & 0 & 0 & 0 & 0 & 0 \\ 0 & 0 & 0 & 0 & 0 & -\xi & 0 & 0 \\ 0 & 0 & 0 & 0 & 0 & 0 & 1 & 0 \\ 0 & 0 & 0 & \xi & 0 & 0 & 0 & 0 \\ 0 & 0 & 0 & 0 & 1 & 0 & 0 & 0 \\ 0 & -\xi & 0 & 0 & 0 & 0 & 0 & 0 \end{pmatrix}, \quad \hat{k} = \begin{pmatrix} 0 & 0 & 0 & 1 & 0 & 0 & 0 & 0 \\ 0 & 0 & 0 & 0 & 0 & 0 & -\xi & 0 \\ 0 & 0 & 0 & 0 & 0 & \xi & 0 & 0 \\ 1 & 0 & 0 & 0 & 0 & 0 & 0 & 0 \\ 0 & 0 & 0 & 0 & 0 & 0 & 0 & 1 \\ 0 & 0 & -\xi & 0 & 0 & 0 & 0 & 0 \\ 0 & \xi & 0 & 0 & 0 & 0 & 0 & 0 \\ 0 & 0 & 0 & 0 & 1 & 0 & 0 & 0 \end{pmatrix},$$

$$\hat{I} = \begin{pmatrix} 0 & 0 & 0 & 0 & 0 & 1 & 0 & 0 \\ 0 & 0 & 0 & 0 & 1 & 0 & 0 & 0 \\ 0 & 0 & 0 & \xi & 0 & 0 & 0 & 0 \\ 0 & 0 & -\xi & 0 & 0 & 0 & 0 & 0 \\ 0 & 1 & 0 & 0 & 0 & 0 & 0 & 0 \\ 1 & 0 & 0 & 0 & 0 & 0 & 0 & 0 \\ 0 & 0 & 0 & 0 & 0 & 0 & 0 & \xi \\ 0 & 0 & 0 & 0 & 0 & 0 & -\xi & 0 \end{pmatrix}, \quad \hat{J} = \begin{pmatrix} 0 & 0 & 0 & 0 & 0 & 0 & 1 & 0 \\ 0 & 0 & 0 & -\xi & 0 & 0 & 0 & 0 \\ 0 & 0 & 0 & 0 & 1 & 0 & 0 & 0 \\ 0 & \xi & 0 & 0 & 0 & 0 & 0 & 0 \\ 0 & 0 & 1 & 0 & 0 & 0 & 0 & 0 \\ 0 & 0 & 0 & 0 & 0 & 0 & 0 & -\xi \\ 1 & 0 & 0 & 0 & 0 & 0 & 0 & 0 \\ 0 & 0 & 0 & 0 & 0 & \xi & 0 & 0 \end{pmatrix}, \quad \hat{K} = \begin{pmatrix} 0 & 0 & 0 & 0 & 0 & 0 & 0 & 1 \\ 0 & 0 & \xi & 0 & 0 & 0 & 0 & 0 \\ 0 & -\xi & 0 & 0 & 0 & 0 & 0 & 0 \\ 0 & 0 & 0 & 0 & 1 & 0 & 0 & 0 \\ 0 & 0 & 0 & 1 & 0 & 0 & 0 & 0 \\ 0 & 0 & 0 & 0 & 0 & 0 & \xi & 0 \\ 0 & 0 & 0 & 0 & 0 & -\xi & 0 & 0 \\ 1 & 0 & 0 & 0 & 0 & 0 & 0 & 0 \end{pmatrix},$$

$$\hat{E} = \begin{pmatrix} 0 & 0 & 0 & 0 & 1 & 0 & 0 & 0 \\ 0 & 0 & 0 & 0 & 0 & 1 & 0 & 0 \\ 0 & 0 & 0 & 0 & 0 & 0 & 1 & 0 \\ 0 & 0 & 0 & 0 & 0 & 0 & 0 & 1 \\ 1 & 0 & 0 & 0 & 0 & 0 & 0 & 0 \\ 0 & 1 & 0 & 0 & 0 & 0 & 0 & 0 \\ 0 & 0 & 1 & 0 & 0 & 0 & 0 & 0 \\ 0 & 0 & 0 & 1 & 0 & 0 & 0 & 0 \end{pmatrix}.$$

**Appendix 2. The simplest eigenfunctions of octonic operators**

All operators of octonic basis have two eigenvalues $\lambda = \pm 1$, which are fourthly degenerate. The simplest eigenfunctions are represented in the table 3.

*Table 3. The simplest eigenfunctions of spatial octonic operators.*

|  | $\lambda = +1$ | | | | $\lambda = -1$ | | | |
|---|---|---|---|---|---|---|---|---|
| $\hat{i}$ | $(1+i)$ | $(j+\xi K)$ | $(E+I)$ | $(J+\xi k)$ | $(1-i)$ | $(j-\xi K)$ | $(E-I)$ | $(J-\xi k)$ |
| $\hat{j}$ | $(1+j)$ | $(k+\xi I)$ | $(E+J)$ | $(K+\xi i)$ | $(1-j)$ | $(k-\xi I)$ | $(E-J)$ | $(K-\xi i)$ |
| $\hat{k}$ | $(1+k)$ | $(i+\xi J)$ | $(E+K)$ | $(I+\xi j)$ | $(1-k)$ | $(i-\xi J)$ | $(E-K)$ | $(I-\xi j)$ |
| $\hat{I}$ | $(1+I)$ | $(J+\xi K)$ | $(E+i)$ | $(j+\xi k)$ | $(1-I)$ | $(J-\xi K)$ | $(E-i)$ | $(j-\xi k)$ |
| $\hat{J}$ | $(1+J)$ | $(K+\xi I)$ | $(E+j)$ | $(k+\xi i)$ | $(1-J)$ | $(K-\xi I)$ | $(E-j)$ | $(k-\xi i)$ |
| $\hat{K}$ | $(1+K)$ | $(I+\xi J)$ | $(E+k)$ | $(i+\xi j)$ | $(1-K)$ | $(I-\xi J)$ | $(E-k)$ | $(i-\xi j)$ |
| $\hat{E}$ | $(1+E)$ | $(i+I)$ | $(j+J)$ | $(k+K)$ | $(1-E)$ | $(i-I)$ | $(j-J)$ | $(k-K)$ |
| $\hat{R}$ | $1$ | $I$ | $J$ | $K$ | $E$ | $i$ | $j$ | $k$ |

**Appendix 3. The multiplication rules for the simplest eigenfunctions of the operator $\hat{K}$**

The simplest eigenfunctions of the operator $\hat{K}$ generate the closed algebra. The rules of multiplication for the functions containing scalar and pseudovector parts are represented in the table 4.

*Table 4. The rules of multiplication for the simplest eigenfunctions of the operator $\hat{K}$.*

|  | $\dfrac{1+K}{2}$ | $\dfrac{1-K}{2}$ | $\dfrac{I+\xi J}{2}$ | $\dfrac{I-\xi J}{2}$ |
|---|---|---|---|---|
| $\dfrac{1+K}{2}$ | $\dfrac{1+K}{2}$ | $0$ | $\dfrac{I+\xi J}{2}$ | $0$ |
| $\dfrac{1-K}{2}$ | $0$ | $\dfrac{1-K}{2}$ | $0$ | $\dfrac{I-\xi J}{2}$ |
| $\dfrac{I+\xi J}{2}$ | $0$ | $\dfrac{I+\xi J}{2}$ | $0$ | $\dfrac{1+K}{2}$ |
| $\dfrac{I-\xi J}{2}$ | $\dfrac{I-\xi J}{2}$ | $0$ | $\dfrac{1-K}{2}$ | $0$ |

**Appendix 4. Algebra of operators of discrete turns and reflections**

The operators of reflection and discrete turns on the angle $\pi$ generate closed algebra. The rules of multiplication for these operators are represented in the table 5.

*Table 5. The rules of multiplication for operators of reflection and discrete turns on the angle $\pi$.*

|            | $\hat{R}_x$ | $\hat{R}_y$ | $\hat{R}_z$ | $\hat{R}$   | $\hat{\pi}_x$ | $\hat{\pi}_y$ | $\hat{\pi}_z$ |
|------------|-------------|-------------|-------------|-------------|---------------|---------------|---------------|
| $\hat{R}_x$ | **1**       | $\hat{\pi}_z$ | $\hat{\pi}_y$ | $\hat{\pi}_x$ | $\hat{R}$     | $\hat{R}_z$   | $\hat{R}_y$   |
| $\hat{R}_y$ | $\hat{\pi}_z$ | **1**       | $\hat{\pi}_x$ | $\hat{\pi}_y$ | $\hat{R}_z$   | $\hat{R}$     | $\hat{R}_x$   |
| $\hat{R}_z$ | $\hat{\pi}_y$ | $\hat{\pi}_x$ | **1**       | $\hat{\pi}_z$ | $\hat{R}_y$   | $\hat{R}_x$   | $\hat{R}$     |
| $\hat{R}$   | $\hat{\pi}_x$ | $\hat{\pi}_y$ | $\hat{\pi}_z$ | **1**       | $\hat{R}_x$   | $\hat{R}_y$   | $\hat{R}_z$   |
| $\hat{\pi}_x$ | $\hat{R}$   | $\hat{R}_z$   | $\hat{R}_y$   | $\hat{R}_x$   | **1**         | $\hat{\pi}_z$ | $\hat{\pi}_y$ |
| $\hat{\pi}_y$ | $\hat{R}_z$ | $\hat{R}$   | $\hat{R}_x$   | $\hat{R}_y$   | $\hat{\pi}_z$ | **1**         | $\hat{\pi}_x$ |
| $\hat{\pi}_z$ | $\hat{R}_y$ | $\hat{R}_x$ | $\hat{R}$   | $\hat{R}_z$   | $\hat{\pi}_y$ | $\hat{\pi}_x$ | **1**         |

**Appendix 5. The rules for commutation of octonic basis operators with operators of discrete turns and reflections**

The operators of discrete turns on the angle $\pi$ and reflection operators either commute ("+") or anticommute ("-") with basis operators that indicated in the table 6.

*Table 6. The rules for commutation of octonic basis operators with operators of discrete turns and reflections.*

|              | $\hat{i}$ | $\hat{j}$ | $\hat{k}$ | $\hat{E}$ | $\hat{I}$ | $\hat{J}$ | $\hat{K}$ |
|--------------|-----------|-----------|-----------|-----------|-----------|-----------|-----------|
| $\hat{R}_x$  | −         | +         | +         | −         | +         | −         | −         |
| $\hat{R}_y$  | +         | −         | +         | −         | −         | +         | −         |
| $\hat{R}_z$  | +         | +         | −         | −         | −         | −         | +         |
| $\hat{R}$    | −         | −         | −         | −         | +         | +         | +         |
| $\hat{\pi}_x$ | +         | −         | −         | +         | +         | −         | −         |
| $\hat{\pi}_y$ | −         | +         | −         | +         | −         | +         | −         |
| $\hat{\pi}_z$ | −         | −         | +         | +         | −         | −         | +         |

**Appendix 6. The spatial rotation of the octonic wave function**

The scalar and pseudoscalar parts of the octonic wave function $\breve{\psi}$ are not transformed under spatial rotation. Therefore we consider the turn of vector part $\vec{\psi}$ on angle $\theta$ round spatial axis defined by the unit pseudovector $\vec{n}$ (see Fig. 1 and 2)

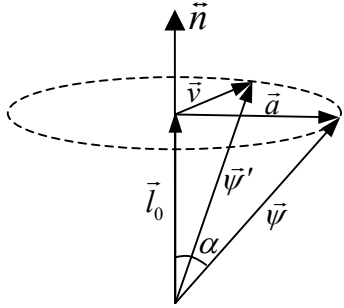
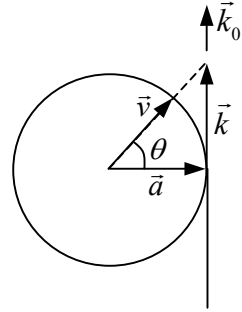

Fig. 1.                           Fig. 2.

Let us consider the polar vector $\vec{l}_0 = (\vec{n},\vec{\psi})\vec{n}$ (see Fig. 1) and vector $\vec{a} = \vec{\psi} - \vec{l}_0$. Using the rules of octonic multiplication the vector $\vec{a}$ can be represented as

$$\vec{a} = \vec{\psi} - (\vec{n},\vec{\psi})\vec{n} = [\vec{n},[\vec{n},\vec{\psi}]].$$

Let $\alpha$ is the angle between vectors $\vec{\psi}$ and $\vec{l}_0$. Then $|\vec{a}| = |\vec{\psi}|\sin\alpha$. Let $\vec{v}$ is vector obtained by rotation of vector $\vec{a}$ round axis $\vec{n}$ on the angle $\theta$. Then the turned vector $\vec{\psi}'$ is defined by

$$\vec{\psi}' = \vec{l}_0 + \vec{v}.$$

To find vector $\vec{v}$ we will use the geometrical construction shown in Fig. 2. Let $\vec{k}_0$ is the unit vector perpendicular to $\vec{n}$ and $\vec{a}$:

$$\vec{k}_0 = \frac{[\vec{n},\vec{a}]}{\xi|\vec{a}|} = -\xi\frac{[\vec{n},\vec{a}]}{|\vec{a}|} = -\xi\frac{[\vec{n},\vec{\psi}]}{|\vec{\psi}|\sin\alpha}.$$

Then for the vector $\vec{k}$ (see Fig. 2) we get the following expression:

$$\vec{k} = \vec{k}_0|\vec{a}|tg\,\theta = -\xi\frac{[\vec{n},\vec{\psi}]|\vec{\psi}|\sin\alpha}{|\vec{\psi}|\sin\alpha}tg\,\theta = -\xi[\vec{n},\vec{\psi}]tg\,\theta.$$

At that $|\vec{k}| = |\vec{a}|tg\,\theta$. For the vector $\vec{v}$ we get

$$\vec{v} = \frac{(\vec{a}+\vec{k})}{|\vec{a}+\vec{k}|}|\vec{a}| = (\vec{a}+\vec{k})\frac{|\vec{a}|}{|\vec{a}|}\cos\theta = (\vec{a}+\vec{k})\cos\theta.$$

Thus the expression for the vector $\vec{\psi}'$ has the following form:

$$\vec{\psi}' = (\vec{n},\vec{\psi})\vec{n} + (\vec{a}+\vec{k})\cos\theta = (\vec{n},\vec{\psi})\vec{n} + [\vec{n},[\vec{n},\vec{\psi}]]\cos\theta - \xi[\vec{n},\vec{\psi}]\sin\theta.$$

After algebraic transform we get finally

$$\vec{\psi}' = \vec{\psi}\cos\theta + (\vec{n},\vec{\psi})\vec{n}(1-\cos\theta) - \xi[\vec{n},\vec{\psi}]\sin\theta.$$

Analogous consideration can be made for the turn of pseudovector $\vec{\varphi}$.